\documentclass[%
 reprint,
 amsmath,amssymb,
 aps,
prl
]{revtex4-2}

\usepackage{graphicx}% Include figure files
\usepackage{dcolumn}% Align table columns on decimal point
\usepackage{bm}% bold math
\usepackage{braket}
\usepackage{xcolor}
\usepackage{comment}
\definecolor{crimson}{rgb}{0.7, 0.1, 0.1}
\begin{document}

\preprint{APS/123-QED}

\title{Measuring a localization phase diagram \\ controlled by the interplay of disorder and driving}

\author{Peter Dotti}
 % \email{p\_dotti@ucsb.edu}
\author{Yifei Bai}%
\thanks{Equal contribution.}
\author{Toshihiko Shimasaki}
\thanks{Equal contribution.}
\author{Anna R. Dardia}
\author{David M. Weld}%
\email{weld@ucsb.edu}
\affiliation{%
\textbf{}
Department of Physics, University of California, Santa Barbara, California 93106, USA
}%

% \collaboration{MUSO Collaboration}%\noaffiliation

%  \homepage{http://www.Second.institution.edu/~Charlie.Author}
% \affiliation{
%  Second institution and/or address\\
%  This line break forced% with \\
% }%
% \affiliation{
%  Third institution, the second for Charlie Author
% \affiliation{%
%  Authors' institution and/or address\\
%  This line break forced with \textbackslash\textbackslash
% }%

% \collaboration{CLEO Collaboration}%\noaffiliation

\date{\today}% It is always \today, today,

\begin{abstract}
The interplay of various localizing mechanisms
is a central topic of modern condensed matter physics.
In this work we experimentally explore the interplay between quasiperiodic disorder and periodic driving, each of which in isolation is capable of driving a metal-insulator phase transition. Using a 1D quasiperiodic cold-atom chain we measure transport across the full phase diagram varying both drive strength and quasidisorder strength. We observe lobes of metallic phases bounded by quantum phase transitions which depend on both drive and disorder.  While these observations are broadly consistent with expectations from a high-drive-frequency theoretical model, we also observe clear departures from the predictions of this model, including anomalous changes in localization behavior at lower drive frequency. We demonstrate experimentally and theoretically that understanding the full measured phase diagram requires an extension to commonly-used approximate theories of Floquet matter. 
\end{abstract}

\maketitle

{The interplay of localizing phenomena is a topic perennially at the forefront of the study of quantum condensed matter.
In this work, we experimentally and theoretically investigate the interplay of two independent localizing influences: periodic driving and quasiperiodic pseudo-disorder. 
A time-periodic driving force applied at specific amplitudes can cause dynamic localization by flattening out a Floquet band~\cite{holthaus-dynloctheory_DL}. 
Separately, quasiperiodic pseudo-disorder can drive a localization-delocalization phase transition via Anderson-insulator-like interference effects, as expressed in the 1D Aubry-Andr\'e-Harper (AAH) model~\cite{aubry1980analyticity,Roati2008Natur.453..895R,lahini_observation_2009_AAExp, Moratti_Modugno_2012_interference, rayanov_decohering_2013}.   
This work aims to explore open questions concerning the {simultaneous} application of \emph{both} of these localizing influences to a quantum system: {how} does the driving {affect} the Anderson-like insulator? {Does} quasidisorder disrupt dynamic localization? What phase diagram arises from the interplay of the two effects?  {How does the phase diagram depend on drive frequency? Such questions are important not only for applications in novel Floquet engineering but also for expanding our understanding of light-induced dynamical control of solids.}}

\begin{figure}[th!]
\includegraphics{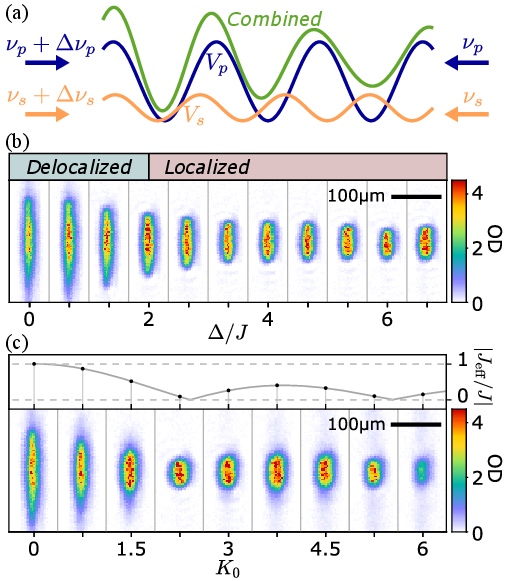}% Here is how to import EPS art
\caption{Observing two kinds of localization. \label{fig:intro_fig} (a)~Primary (blue) and secondary (orange) lattices combine to create a quasiperiodic potential (green) which can be translated via frequency modulation.  (b)~Atoms imaged after 500~ms hold time  for different secondary lattice strengths demonstrate AAH localization at $\Delta/J>2$.  (c)~Atoms imaged after 500~ms hold time  for different drive strengths $K_0$ demonstrate dynamic localization.  Top panel indicates theoretically predicted $|J_\mathrm{eff}|$. 
The lattice wavevector is in the vertical direction in (b) and (c).}
\end{figure}

Previous work has theoretically explored transport in a driven disordered  lattice~ \cite{Holthaus_Hone_DL+Anderson_95,Holthaus_Hone_DL+Anderson_96,Drese+Holthaus_97,HolthausChapter_DL+AAH} and experimentally investigated the effects of disorder in the vicinity of dynamic localization in waveguides~\cite{Guzman-Silva:20}.  These studies were generally restricted to the high-frequency limit.

The experiments begin by loading a Bose-Einstein condensate of $8 \times 10^4$ $^{84}$Sr atoms into a pair of superimposed 1D optical lattices with different spatial periods aligned along an axis perpendicular to gravity.  The primary (secondary) lattice is formed by two counter-propagating beams of wavelength $\lambda_p=1064$~nm ($\lambda_s=874.61$~nm $\equiv\lambda_p/\beta$). All four beams pass through individual acousto-optic modulators. This allows for control of the lattice depths, denoted $V_p$ and $V_s$ for the primary and secondary lattices, and for controlled translation of the primary (secondary) lattice by adjusting the frequency difference $\Delta \nu_{p}$ ($\Delta \nu_{s}$) between the two beams forming the lattice, which both have frequency $\nu_p$ ($\nu_s$) otherwise~(Fig.~\ref{fig:intro_fig}(a)). The resulting ability to rapidly tune the depth and spatial phase of both lattice potentials enables the simultaneous experimental realization of dynamic and quasidisorder-induced localization.
More experimental details appear in~\cite{supp_mat}.

Synchronized sinusoidal translation of both lattices is conveniently treated by transforming to the co-moving frame, in which the modulation appears as a time-varying electric field. We focus on the case where the secondary lattice is at rest in the co-moving frame of the primary lattice, implemented by fixing $\Delta \nu_{s} = \beta \Delta \nu_{p}$. 
This admits a description by the tight-binding Hamiltonian~\cite{Drese+Holthaus_97}
\begin{equation}
\begin{split}
H =& \sum_{l}-J\left(\ket{l+1}\!\bra{l} +\mathrm{h.c.} \right) + \Delta \cos(2\pi\beta l-\delta)\ket{l}\!\bra{l} \\
    &+ \hbar \omega K_0 \cos(\omega t) l \ket{l}\!\bra{l},
\label{basicHamiltonian_TB}
\end{split}
\end{equation}
where $J$ is the tunneling energy, $\Delta$ is the quasidisorder strength, $K_0$ is the dimensionless drive strength, $\omega$ is the angular drive frequency, $\beta = \lambda_p/\lambda_s=1064/874.61$, and the phason parameter $\delta$ is constant.  {The tight-binding approximation is justified when $V_p$ and $V_p/V_s$ are sufficiently large that only nearest-neighbor tunneling contributes significantly, as is the case in our experiments~\cite{Li+Das_Sarma_2017,Modugno_2009}.  $V_p$ is chosen to be 9$E_{R,p}$, leading to $J=0.024E_{R,p} = h \times 50\text{~Hz}$,  where $E_{R,p} = 2\pi^2\hbar^2/m\lambda_p^2$ is the recoil energy of the primary lattice. We explore a range of $V_s$ values, from 0 up to 0.59$E_{R,p}$. Confinement transverse to the lattice axis allows the state to be well approximated as a product of a 1D component obeying equation \eqref{basicHamiltonian_TB} and a static 2D transverse state \cite{supp_mat}.} 

Two special cases of Hamiltonian~\eqref{basicHamiltonian_TB} correspond to the two localization mechanisms previously mentioned. When $K_0 = 0$, the Hamiltonian becomes the static AAH model. For irrational $\beta$~\cite{Jitomirskaya-AAH-Mathematics}, all eigenstates of this Hamiltonian exhibit a localization-delocalization quantum phase transition at $\Delta/J=2$. Fig.~\ref{fig:intro_fig}(b) experimentally demonstrates the phase transition of the AAH model, showing \emph{in situ} absorption images taken after 500~ms of expansion in the bichromatic lattice potential for $K_0=0$ with $\Delta/J$ values ranging from 0 to 6.7.  As in Ref.~\cite{Roati2008Natur.453..895R}, the phase transition is observed in the suppression of expansion for values of $\Delta/J>2$. {We note that the existence of a mean-field interaction causes subdiffusion~\cite{Lucioni_subdiff_PhysRevLett.106.230403, Deissler2010}, signaling an insulating phase in the non-interacting limit.}  If instead $\Delta = 0$ in Hamiltonian (1), dynamic localization can occur~\cite{eckardt_superfluid-insulator_2005,Zoller_DL_92,Raizen_DL_98,arimondo-shakenlattice}. Applying Floquet theory~\cite{holthaus_floquet_2016,Floquet_reform_with_avg_E_22} results in a modified tight-binding Hamiltonian with effective tunneling energy $J_\mathrm{eff} = J \mathcal{J}_0(K_0)$~\cite{eckardt_superfluid-insulator_2005}. Crucially, hopping is entirely suppressed for dimensionless drive amplitudes corresponding to zeros of the Bessel function $\mathcal{J}_0$, for example at $K_0\approx 2.405$ and $5.520$. Fig.~\ref{fig:intro_fig}(c) shows \emph{in situ} absorption images taken after 500~ms of expansion for varying drive amplitudes at $\Delta=0$ and $\omega = 2\pi\times(1\mathrm{\ kHz})= 20J/\hbar$. Expansion is initiated by sudden removal of a confining optical dipole trap, simultaneous with the beginning of sinusoidal phase modulation. These data clearly demonstrate the suppression of transport due to dynamic localization for $K_0$ close to the zeroes of $\mathcal{J}_0(K_0)$.

%%%%%%%%%%%%%%%%%%%%%%%%%%%%%%%%
%%%%% Figure 2 Phase Diagram %%%%%
\begin{figure*}[t]
\includegraphics{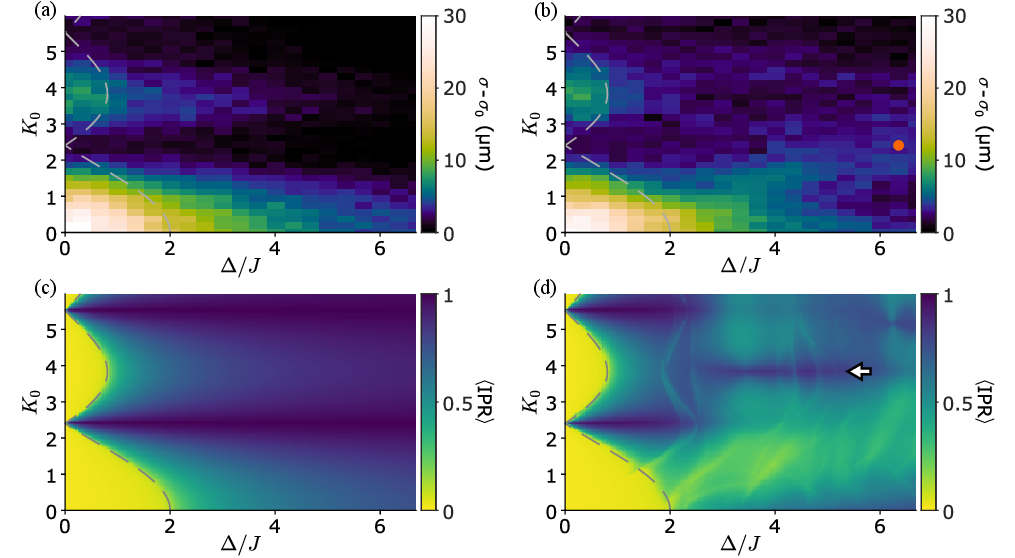}% Here is how to import EPS art
\caption{\label{fig:phaseDiagram} Interplay of dynamic localization and quasidisorder-induced localization in two different drive frequency regimes. (a)~Measured width $\sigma$ (standard deviation of a Gaussian fit) of the atomic density distribution after a 500~ms hold minus its initial value $\sigma_0=10$~\textmu m, for a high drive frequency $\omega = 2\pi \times(1\mathrm{\ kHz}) = 20J/\hbar$.
(b) Same quantity for a lower drive frequency $\omega = 2\pi \times(200\mathrm{\ Hz}) = 3.93J/\hbar$, near the width of the ground band. The orange dot marks the experimental condition in Fig. \ref{fig:expansionInLocalizedRegime}. (c) and (d) show numerically calculated average Floquet eigenstate IPR for the 1~kHz drive and the 200~Hz drive respectively. In all panels dashed lines indicate phase boundaries predicted by the approximate model of Eq.~\eqref{basicHamiltonian_TB_FL}.  In the region $\Delta/J>2$, we observe clear differences between (a) and (b) and between (c) and (d), indicating differing Floquet eigenstate properties in the two regimes of drive frequency. $V_p=9E_{R,p}$ for all panels.  The white arrow in (d) indicates the observed region of enhanced localization, and corresponds to the arrow in Fig.~\ref{fig:FrequenyComaparison}(a).}
\end{figure*}

Having verified that the experimental system can exhibit both quasidisorder-induced localization and drive-induced localization, we proceed to investigate the main topic of this work, which is the interplay of these two phenomena. We measured the width of the atomic density distribution after 500~ms of expansion, with quasidisorder strength $\Delta$ and drive amplitude $K_0$ independently varied to map out a 2D parameter space. Figs.~\ref{fig:phaseDiagram}(a)~and~(b) present experimental measurements of the localization phase diagram for a drive frequency of 1~kHz ($\hbar\omega = 20J$) and 200~Hz ($\hbar\omega = 3.9J$), respectively.   For comparison, Figs.~\ref{fig:phaseDiagram}(c) and \ref{fig:phaseDiagram}(d) present numerically determined average inverse participation ratio $\langle\mathrm{IPR}\rangle$, calculated by averaging the IPR values of the Floquet eigenstates as detailed in \cite{supp_mat}.  

The basic structure of Figs.~2(a) and (c) {can be understood by applying the unitary transformation $\hat U = \exp\left\{ -i K_0 \sin(\omega t) \sum_l l\ket{l}\!\bra{l}\right \}$ to Hamiltonian~\eqref{basicHamiltonian_TB}. This converts the oscillating linear force to an oscillating phase in the tunneling strength \cite{eckardt_superfluid-insulator_2005, Tarallo_PhysRevA.86.033615, MiyakePRL}: }  

\begin{equation}
\begin{split}
H(t)\rightarrow H'(t) =& \sum_{l}-J\left(e^{iK_0 \sin(\omega t)}\ket{l+1}\!\bra{l} +\mathrm{h.c.} \right) \\
&+ \Delta \cos(2\pi\beta l-\delta)\ket{l}\!\bra{l}.
\label{basicHamiltonian_TB_transformed}
\end{split}
\end{equation}
{With the high-frequency approximation} ($\hbar\omega \gg 4J, 2\Delta$){, the dynamics can be well-captured by the time-averaged Hamiltonian}:
\begin{equation}
\begin{split}
H_\text{eff} \approx& \sum_{l} -J_\mathrm{eff}\left(\ket{l+1}\!\bra{l} +\mathrm{h.c.} \right)+ \Delta \cos(2\pi\beta l-\delta)\ket{l}\!\bra{l}
\label{basicHamiltonian_TB_FL}
\end{split}
\end{equation}
Eq.~(2) is the AAH model with a Bessel-renormalized hopping amplitude $J_\mathrm{eff}$. Thus, the localization phase transition occurs at $\Delta_c = 2|J_\mathrm{eff}| = 2|J\mathcal{J}_0 (K_0)|$. This theoretically predicted phase boundary is indicated by dashed curves in Fig.~\ref{fig:phaseDiagram}. 

The first main result of this work is the experimental observation of the predicted delocalized lobes in the region $|\Delta/J_\mathrm{eff}|<2$. The data clearly exhibit a drive-amplitude-dependent localization-delocalization phase transition, featuring metallic lobes separated by fingers of an insulating phase touching the $\Delta=0$ axis at the zeroes of the Bessel function $\mathcal{J}_0(K_0)$ where $J_\mathrm{eff}=0$. The analytic $\Delta/|J_\mathrm{eff}|=2$ curves correspond very well to the transition observed both in experimental expansion data (Figs.~\ref{fig:phaseDiagram}(a) and (b)) and in numerically calculated $\langle\mathrm{IPR}\rangle$ (Figs.~\ref{fig:phaseDiagram}(c) and (d)).
The transition between phases is intrinsically somewhat less sharply seen in measured change in width $\sigma - \sigma_0$ than $\langle\mathrm{IPR}\rangle$ because $\sigma - \sigma_0$ scales with the effective hopping time $\hbar/|J_\mathrm{eff}|$ even in delocalized regions. The form of the measured high-frequency phase diagram of Fig.~\ref{fig:phaseDiagram}(a) agrees with both the numerical theory of Fig. \ref{fig:phaseDiagram}(b) and the analytic approximation of Eq.~\eqref{basicHamiltonian_TB_FL}, indicating that these data constitute observation of the phase boundaries arising from the interplay between quasidisorder and driving. 

\begin{figure}[th!]
\includegraphics{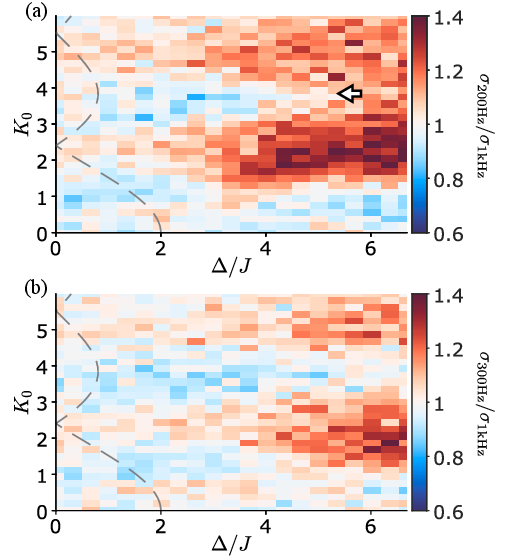}
\caption{\label{fig:FrequenyComaparison} Measured ratio between high and low-frequency phase diagrams. (a)~Ratio between the expansion with drive frequency $\omega = 2\pi \times(200\mathrm{\ Hz}) = 3.93J/\hbar$ and $\omega = 2\pi \times(1\mathrm{\ kHz}) = 20J/\hbar$.  The data used to generate (a) is the same as that presented in Fig.~\ref{fig:phaseDiagram}. (b)~Ratio between data taken at drive frequency $\omega = 2\pi \times(300\mathrm{\ Hz}) = 5.90J/\hbar$ and the same 1~kHz data.  The white arrow in (a) is the same as that in Fig.~\ref{fig:phaseDiagram}(d) and indicates the observed region of enhanced localization.}
\end{figure}

For lower drive frequencies, comparable to the bandwidth, new features appear in both the experimental and numerical phase diagrams which are not captured by the model of Eq.\!~\eqref{basicHamiltonian_TB_FL}. The observation and characterization of these new features constitutes the second main result of this work. These differences are apparent in Figs.~\ref{fig:phaseDiagram}(b) and \ref{fig:phaseDiagram}(d). We visualize the differences more starkly in Fig.~{\ref{fig:FrequenyComaparison}(a)}, which shows the ratio between the measured low-frequency and high-frequency widths, denoted $\sigma_{200\mathrm{Hz}}/\sigma_{1\mathrm{kHz}}$.  We observe the dominant deviation from the high-frequency behavior to occur when $\Delta/J\gtrsim 2$.  We note in particular two features, which appear in both the experimental data of Fig.~\ref{fig:phaseDiagram}(b) and \ref{fig:FrequenyComaparison}(a) and the numerical results of Fig.~\ref{fig:phaseDiagram}(d):  a narrow band of anomalously low expansion near the line $K_0\approx 3.8$ for larger values of $\Delta/J$, and a broad area of anomalously high expansion in the lower-right of the parameter space. We interpret these features as a signature of nontrivial phenomena not captured by the simple high-frequency approximation.

To further investigate the drive frequency dependence of these features of the phase diagram, we took similar measurements of the expansion with a drive frequency of 300~Hz ($\hbar\omega = 5.90J$), but otherwise identical experimental conditions, presented in~\cite{supp_mat}. The ratio between the measured widths at 300~Hz and the 1~kHz data set, denoted $\sigma_{300\mathrm{Hz}}/\sigma_{1\mathrm{kHz}}$, appears in Fig.~\ref{fig:FrequenyComaparison}(b). We observe the same general features at this slightly higher drive frequency, but the extent and magnitude of the deviation is reduced, indicating the approach towards the high-frequency limit.

The first anomalous feature of the low-drive-frequency phase diagram, the observed enhanced localization near $K_0 \approx 3.8$, corresponds closely to a zero of the \emph{first} Bessel function $\mathcal{J}_1(K_0)$. In the high-frequency limit, the nonzero-frequency Bessel terms average away over many cycles. However, when $\hbar \omega \approx 4J$, higher-order terms {from the Jacobi-Anger expansion} contribute,
\[
e^{iK_0 \sin(\omega t)} = \sum_{m=-\infty}^\infty \mathcal{J}_m(K_0) e^{im\omega t},
\]
with the $\mathcal{J}_1(K_0)$ term the most relevant term past leading order.  We thus interpret the observed localized stripe near $K_0\approx 3.8$, indistinguishable within experimental error from the first zero of the $\mathcal{J}_1$ Bessel function near 3.8317, as the effect of a beyond-leading-order localization process not captured in the commonly-used high-frequency approximate description of dynamic localization.

%%%%%%%%%%%%%%%%%%%%%%%%%
%%%%% FIG 4 DROP %%%%%
\begin{figure}[th!]
\includegraphics[width=3.375in]{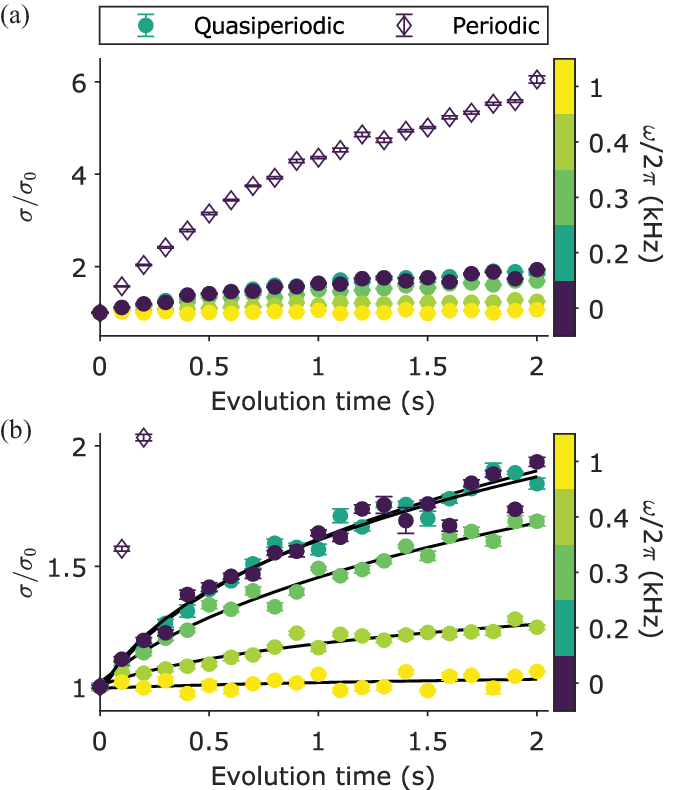}% Here is how to import EPS art
\caption{\label{fig:expansionInLocalizedRegime} \label{fig:timeExpansion_Version_b} Measured expansion versus time in various regimes. (a) Expansion in a purely periodic lattice (diamond markers) is much more rapid than that in the quasiperiodic lattice in the localized phase (circular markers). (b) Same data as in (a) with y axis expanded to show variation among the expansions in quasiperiodic potentials. Different driving frequencies are represented by the color of the marker. Each data point is averaged from 5 experimental realizations. Here $\Delta/J = 6.35$ and $K_0 \approx 2.4046$. Error bars show standard error of the mean.}
\end{figure}

%%%%%%%%%%%%%%%%%%%%%%%%%
%%%%% table for transport exponent data %%%%%
\begin{table}[th!]
\begin{tabular}{|c|c|c|c|c|c|}
\hline
$\omega/2\pi \, (\mathrm{kHz})$ & 0       & 0.2     & 0.3     & 0.4     & 1        \\ \hline
$\gamma$                       & 0.24(5) & 0.26(5) & 0.26(7) & 0.12(6) & 0.03(21) \\ \hline
\end{tabular}
\label{table:transportExp}
\caption{Transport exponents $\gamma$ extracted by fitting the expansion of the condensate in a quasiperiodic lattice under various driving frequencies. Error bars represent 95\% confidence interval. }
\end{table}

The second anomalous feature, the apparent decrease in localization in the lower-right quadrant of the low-drive-frequency phase diagram, requires a separate explanation. This feature could in principle arise from true delocalization of the Floquet eigenstates or from the existence of an extended critical phase.  However, as discussed below {and in the Supplementary Material \cite{supp_mat}}, additional measurements of transport dynamics and numerical IPR calculations indicate that this region remains a localized phase in the presence of driving, but with an increased localization length. 
 
To explore the nature of this region of the phase diagrams, we choose the representative point $\Delta/J \approx 6.35$ (a strong quasiperiodic potential) and $K_0 \approx 2.4046$.  The considered potential alone in the absence of driving and interactions is well in the localized phase of the AAH model; the considered drive alone essentially produces dynamic localization ($\mathcal{J}_0(K_0) = 1.3\times 10^{-4}$) in the absence of quasiperiodicity. To experimentally investigate the interplay between quasidisorder-induced and dynamic localization at various drive frequencies, we conduct transport measurements  observing the evolution of the width of the condensate for up to 2 seconds ($\approx 639.4\hbar/J$). The results are shown in Fig.~\ref{fig:expansionInLocalizedRegime}, along with measurements of the typical fast expansion in a static periodic lattice ($\Delta=0$), shown with diamond markers in Fig.~\ref{fig:expansionInLocalizedRegime}(a) as a point of comparison.

We observe in {the} data of Fig.~\ref{fig:expansionInLocalizedRegime}(b) that when the driving frequency is low, as for $\omega = 2\pi\times 200 \,\mathrm{Hz}$, the measured evolution follows that of the static quasiperiodic system (denoted as $\omega=0$). The expansion slows further as the driving frequency increases. Notably, the expansion is totally arrested for the duration of the experiment in the high-frequency case ($\omega = 2\pi\times 1\,\mathrm{kHz}$).  In all  cases reported, the measured expansion is slower than for the localized phase of the static quasiperiodic case.
{We extract a transport exponent $\gamma$ for each drive frequency from the data in Fig.~\ref{fig:expansionInLocalizedRegime} and present it in Table~I.  We find that the expansion is sub-diffusive as in the static quasiperiodic case, and is reduced at high drive frequencies. These observations suggest that this anomalous region is still a localized phase.}

{The experimental results are supported by calculations of the scaling of IPR versus system size, which indicate the phase to be fully localized. Numerical calculations of the IPR indicate that this anomalous region is distinguished by larger localization lengths which are more strongly dependent on quasienergy than in the high-frequency case. This is in stark contrast to the static AAH model or the high-frequency approximation of the driven case, in which the localization lengths of the eigenstates, and thus their IPR, are independent of the (quasi)energy. Further details on the experimental and numerical analysis are provided in the Supplementary Information \cite{supp_mat}}.
Therefore, although the expansion of the condensate in the low-frequency regime resembles that of the static quasiperiodic lattice, their mechanisms are completely different. 

%%%%%%%%%%%%%%%%%%%%%%%%%%%%%%%%%%
%%%%% Discussion and Outlook %%%%%

In summary, we have experimentally and theoretically investigated the interplay of dynamic and quasidisorder-induced localization. In the high-frequency regime, the measured phase diagram is consistent with both numerical and time-averaged analytic theory, featuring metallic Bessel lobes separated from insulating regions by a phase boundary which depends on both drive strength and quasidisorder strength.
For lower frequency driving, comparable to the bandwidth, we observe localization behavior not captured by the simple time-averaged analytic model, including areas of both enhanced and suppressed transport, and show that it can be understood via extensions to {this} simple {high-frequency} model. 
This work touches on some of the rich structures that can arise when quantum systems are driven at frequencies near their natural energy scales, which may prove relevant for {both broadening the applicability of Floquet engineering techniques and for understanding results of optically driven solid-state experiments}. A natural direction for future work is the addition of tunable interactions, enabling experimental probes of the interplay among three distinct fundamental localizing mechanisms of many-body quantum matter: interactions, driving, and disorder~\cite{Bairey_PhysRevB.96.020201}.

%\textbf{\emph{Acknowledgements}\\
\begin{acknowledgments}
We acknowledge inspirational conversations with Martin Holthaus, and research  support from the Air Force Office of Scientific Research (FA9550-20-1-0240) and the NSF QLCI program (OMA-2016245). A.R.D. acknowledges support from the UC Santa Barbara NSF Quantum Foundry funded via the Q-AMASE-i program under Grant DMR1906325. The optical lattices used herein were developed in work supported by the U.S. Department of Energy, Office of Science, National Quantum Information Science Research Centers, Quantum Science Center.
\end{acknowledgments}

\bibliography{mainBib}% Produces the bibliography via BibTeX.

\begin{thebibliography}{33}%
\makeatletter
\providecommand \@ifxundefined [1]{%
 \@ifx{#1\undefined}
}%
\providecommand \@ifnum [1]{%
 \ifnum #1\expandafter \@firstoftwo
 \else \expandafter \@secondoftwo
 \fi
}%
\providecommand \@ifx [1]{%
 \ifx #1\expandafter \@firstoftwo
 \else \expandafter \@secondoftwo
 \fi
}%
\providecommand \natexlab [1]{#1}%
\providecommand \enquote  [1]{``#1''}%
\providecommand \bibnamefont  [1]{#1}%
\providecommand \bibfnamefont [1]{#1}%
\providecommand \citenamefont [1]{#1}%
\providecommand \href@noop [0]{\@secondoftwo}%
\providecommand \href [0]{\begingroup \@sanitize@url \@href}%
\providecommand \@href[1]{\@@startlink{#1}\@@href}%
\providecommand \@@href[1]{\endgroup#1\@@endlink}%
\providecommand \@sanitize@url [0]{\catcode `\\12\catcode `\$12\catcode `\&12\catcode `\#12\catcode `\^12\catcode `\_12\catcode `\%12\relax}%
\providecommand \@@startlink[1]{}%
\providecommand \@@endlink[0]{}%
\providecommand \url  [0]{\begingroup\@sanitize@url \@url }%
\providecommand \@url [1]{\endgroup\@href {#1}{\urlprefix }}%
\providecommand \urlprefix  [0]{URL }%
\providecommand \Eprint [0]{\href }%
\providecommand \doibase [0]{https://doi.org/}%
\providecommand \selectlanguage [0]{\@gobble}%
\providecommand \bibinfo  [0]{\@secondoftwo}%
\providecommand \bibfield  [0]{\@secondoftwo}%
\providecommand \translation [1]{[#1]}%
\providecommand \BibitemOpen [0]{}%
\providecommand \bibitemStop [0]{}%
\providecommand \bibitemNoStop [0]{.\EOS\space}%
\providecommand \EOS [0]{\spacefactor3000\relax}%
\providecommand \BibitemShut  [1]{\csname bibitem#1\endcsname}%
\let\auto@bib@innerbib\@empty
%</preamble>
\bibitem [{\citenamefont {Holthaus}(1992)}]{holthaus-dynloctheory_DL}%
  \BibitemOpen
  \bibfield  {author} {\bibinfo {author} {\bibfnamefont {M.}~\bibnamefont {Holthaus}},\ }\href {https://doi.org/10.1103/PhysRevLett.69.351} {\bibfield  {journal} {\bibinfo  {journal} {Phys. Rev. Lett.}\ }\textbf {\bibinfo {volume} {69}},\ \bibinfo {pages} {351} (\bibinfo {year} {1992})}\BibitemShut {NoStop}%
\bibitem [{\citenamefont {Aubry}\ and\ \citenamefont {Andr{\'e}}(1980)}]{aubry1980analyticity}%
  \BibitemOpen
  \bibfield  {author} {\bibinfo {author} {\bibfnamefont {S.}~\bibnamefont {Aubry}}\ and\ \bibinfo {author} {\bibfnamefont {G.}~\bibnamefont {Andr{\'e}}},\ }\href@noop {} {\bibfield  {journal} {\bibinfo  {journal} {Ann. Israel Phys. Soc.}\ }\textbf {\bibinfo {volume} {3}},\ \bibinfo {pages} {18} (\bibinfo {year} {1980})}\BibitemShut {NoStop}%
\bibitem [{\citenamefont {{Roati}}\ \emph {et~al.}(2008)\citenamefont {{Roati}}, \citenamefont {{D'Errico}}, \citenamefont {{Fallani}}, \citenamefont {{Fattori}}, \citenamefont {{Fort}}, \citenamefont {{Zaccanti}}, \citenamefont {{Modugno}}, \citenamefont {{Modugno}},\ and\ \citenamefont {{Inguscio}}}]{Roati2008Natur.453..895R}%
  \BibitemOpen
  \bibfield  {author} {\bibinfo {author} {\bibfnamefont {G.}~\bibnamefont {{Roati}}}, \bibinfo {author} {\bibfnamefont {C.}~\bibnamefont {{D'Errico}}}, \bibinfo {author} {\bibfnamefont {L.}~\bibnamefont {{Fallani}}}, \bibinfo {author} {\bibfnamefont {M.}~\bibnamefont {{Fattori}}}, \bibinfo {author} {\bibfnamefont {C.}~\bibnamefont {{Fort}}}, \bibinfo {author} {\bibfnamefont {M.}~\bibnamefont {{Zaccanti}}}, \bibinfo {author} {\bibfnamefont {G.}~\bibnamefont {{Modugno}}}, \bibinfo {author} {\bibfnamefont {M.}~\bibnamefont {{Modugno}}},\ and\ \bibinfo {author} {\bibfnamefont {M.}~\bibnamefont {{Inguscio}}},\ }\href {https://doi.org/10.1038/nature07071} {\bibfield  {journal} {\bibinfo  {journal} {\nat}\ }\textbf {\bibinfo {volume} {453}},\ \bibinfo {pages} {895} (\bibinfo {year} {2008})}\BibitemShut {NoStop}%
\bibitem [{\citenamefont {Lahini}\ \emph {et~al.}(2009)\citenamefont {Lahini}, \citenamefont {Pugatch}, \citenamefont {Pozzi}, \citenamefont {Sorel}, \citenamefont {Morandotti}, \citenamefont {Davidson},\ and\ \citenamefont {Silberberg}}]{lahini_observation_2009_AAExp}%
  \BibitemOpen
  \bibfield  {author} {\bibinfo {author} {\bibfnamefont {Y.}~\bibnamefont {Lahini}}, \bibinfo {author} {\bibfnamefont {R.}~\bibnamefont {Pugatch}}, \bibinfo {author} {\bibfnamefont {F.}~\bibnamefont {Pozzi}}, \bibinfo {author} {\bibfnamefont {M.}~\bibnamefont {Sorel}}, \bibinfo {author} {\bibfnamefont {R.}~\bibnamefont {Morandotti}}, \bibinfo {author} {\bibfnamefont {N.}~\bibnamefont {Davidson}},\ and\ \bibinfo {author} {\bibfnamefont {Y.}~\bibnamefont {Silberberg}},\ }\href {https://doi.org/10.1103/PhysRevLett.103.013901} {\bibfield  {journal} {\bibinfo  {journal} {Phys. Rev. Lett.}\ }\textbf {\bibinfo {volume} {103}},\ \bibinfo {pages} {013901} (\bibinfo {year} {2009})}\BibitemShut {NoStop}%
\bibitem [{\citenamefont {Moratti}\ and\ \citenamefont {Modugno}(2012)}]{Moratti_Modugno_2012_interference}%
  \BibitemOpen
  \bibfield  {author} {\bibinfo {author} {\bibfnamefont {M.}~\bibnamefont {Moratti}}\ and\ \bibinfo {author} {\bibfnamefont {M.}~\bibnamefont {Modugno}},\ }\href {https://doi.org/10.1140/epjd/e2012-30132-3} {\bibfield  {journal} {\bibinfo  {journal} {The European Physical Journal D}\ }\textbf {\bibinfo {volume} {66}},\ \bibinfo {pages} {138} (\bibinfo {year} {2012})}\BibitemShut {NoStop}%
\bibitem [{\citenamefont {Rayanov}\ \emph {et~al.}(2013)\citenamefont {Rayanov}, \citenamefont {Radons},\ and\ \citenamefont {Flach}}]{rayanov_decohering_2013}%
  \BibitemOpen
  \bibfield  {author} {\bibinfo {author} {\bibfnamefont {K.}~\bibnamefont {Rayanov}}, \bibinfo {author} {\bibfnamefont {G.}~\bibnamefont {Radons}},\ and\ \bibinfo {author} {\bibfnamefont {S.}~\bibnamefont {Flach}},\ }\href {https://doi.org/10.1103/PhysRevE.88.012901} {\bibfield  {journal} {\bibinfo  {journal} {Phys. Rev. E}\ }\textbf {\bibinfo {volume} {88}},\ \bibinfo {pages} {012901} (\bibinfo {year} {2013})}\BibitemShut {NoStop}%
\bibitem [{\citenamefont {Holthaus}\ \emph {et~al.}(1995)\citenamefont {Holthaus}, \citenamefont {Ristow},\ and\ \citenamefont {Hone}}]{Holthaus_Hone_DL+Anderson_95}%
  \BibitemOpen
  \bibfield  {author} {\bibinfo {author} {\bibfnamefont {M.}~\bibnamefont {Holthaus}}, \bibinfo {author} {\bibfnamefont {G.~H.}\ \bibnamefont {Ristow}},\ and\ \bibinfo {author} {\bibfnamefont {D.~W.}\ \bibnamefont {Hone}},\ }\href {https://doi.org/10.1103/PhysRevLett.75.3914} {\bibfield  {journal} {\bibinfo  {journal} {Phys. Rev. Lett.}\ }\textbf {\bibinfo {volume} {75}},\ \bibinfo {pages} {3914} (\bibinfo {year} {1995})}\BibitemShut {NoStop}%
\bibitem [{\citenamefont {Holthaus}\ and\ \citenamefont {Hone}(1996)}]{Holthaus_Hone_DL+Anderson_96}%
  \BibitemOpen
  \bibfield  {author} {\bibinfo {author} {\bibfnamefont {M.}~\bibnamefont {Holthaus}}\ and\ \bibinfo {author} {\bibfnamefont {D.~W.}\ \bibnamefont {Hone}},\ }\href {https://doi.org/10.1080/01418639608240331} {\bibfield  {journal} {\bibinfo  {journal} {Philos. Mag. B}\ }\textbf {\bibinfo {volume} {74}},\ \bibinfo {pages} {105} (\bibinfo {year} {1996})}\BibitemShut {NoStop}%
\bibitem [{\citenamefont {Drese}\ and\ \citenamefont {Holthaus}(1997)}]{Drese+Holthaus_97}%
  \BibitemOpen
  \bibfield  {author} {\bibinfo {author} {\bibfnamefont {K.}~\bibnamefont {Drese}}\ and\ \bibinfo {author} {\bibfnamefont {M.}~\bibnamefont {Holthaus}},\ }\href {https://doi.org/10.1103/PhysRevLett.78.2932} {\bibfield  {journal} {\bibinfo  {journal} {Phys. Rev. Lett.}\ }\textbf {\bibinfo {volume} {78}},\ \bibinfo {pages} {2932} (\bibinfo {year} {1997})}\BibitemShut {NoStop}%
\bibitem [{\citenamefont {Arlinghaus}\ \emph {et~al.}(2011)\citenamefont {Arlinghaus}, \citenamefont {Langemeyer},\ and\ \citenamefont {Holthaus}}]{HolthausChapter_DL+AAH}%
  \BibitemOpen
  \bibfield  {author} {\bibinfo {author} {\bibfnamefont {S.}~\bibnamefont {Arlinghaus}}, \bibinfo {author} {\bibfnamefont {M.}~\bibnamefont {Langemeyer}},\ and\ \bibinfo {author} {\bibfnamefont {M.}~\bibnamefont {Holthaus}},\ }in\ \href@noop {} {\emph {\bibinfo {booktitle} {Dynamical Tunneling: Theory and Experiment}}},\ \bibinfo {editor} {edited by\ \bibinfo {editor} {\bibfnamefont {S.}~\bibnamefont {Keshavamurthy}}\ and\ \bibinfo {editor} {\bibfnamefont {P.}~\bibnamefont {Schlagheck}}}\ (\bibinfo  {publisher} {CRC Press},\ \bibinfo {year} {2011})\ \bibinfo {edition} {1st}\ ed.,\ Chap.~\bibinfo {chapter} {12}\BibitemShut {NoStop}%
\bibitem [{\citenamefont {Guzman-Silva}\ \emph {et~al.}(2020)\citenamefont {Guzman-Silva}, \citenamefont {Heinrich}, \citenamefont {Biesenthal}, \citenamefont {Kartashov},\ and\ \citenamefont {Szameit}}]{Guzman-Silva:20}%
  \BibitemOpen
  \bibfield  {author} {\bibinfo {author} {\bibfnamefont {D.}~\bibnamefont {Guzman-Silva}}, \bibinfo {author} {\bibfnamefont {M.}~\bibnamefont {Heinrich}}, \bibinfo {author} {\bibfnamefont {T.}~\bibnamefont {Biesenthal}}, \bibinfo {author} {\bibfnamefont {Y.~V.}\ \bibnamefont {Kartashov}},\ and\ \bibinfo {author} {\bibfnamefont {A.}~\bibnamefont {Szameit}},\ }\href {https://doi.org/10.1364/OL.380399} {\bibfield  {journal} {\bibinfo  {journal} {Opt. Lett.}\ }\textbf {\bibinfo {volume} {45}},\ \bibinfo {pages} {415} (\bibinfo {year} {2020})}\BibitemShut {NoStop}%
\bibitem [{sup()}]{supp_mat}%
  \BibitemOpen
  \href@noop {} {}\bibinfo {note} {See Supplementary Material at [URL] for experimental and theoretical details and additional data, which includes Ref.~\cite{shimasaki_anomalous_2024,MLGWS,LucioniPRE_PhysRevE.87.042922,Shimasaki_PhysRevLett.133.083405,Martinez_PhysRevB.73.073104,LocReview_B_Kramer_1993,D_J_Thouless_1972}.}\BibitemShut {Stop}%
\bibitem [{\citenamefont {Li}\ \emph {et~al.}(2017)\citenamefont {Li}, \citenamefont {Li},\ and\ \citenamefont {Das~Sarma}}]{Li+Das_Sarma_2017}%
  \BibitemOpen
  \bibfield  {author} {\bibinfo {author} {\bibfnamefont {X.}~\bibnamefont {Li}}, \bibinfo {author} {\bibfnamefont {X.}~\bibnamefont {Li}},\ and\ \bibinfo {author} {\bibfnamefont {S.}~\bibnamefont {Das~Sarma}},\ }\href {https://doi.org/10.1103/PhysRevB.96.085119} {\bibfield  {journal} {\bibinfo  {journal} {Phys. Rev. B}\ }\textbf {\bibinfo {volume} {96}},\ \bibinfo {pages} {085119} (\bibinfo {year} {2017})}\BibitemShut {NoStop}%
\bibitem [{\citenamefont {Modugno}(2009)}]{Modugno_2009}%
  \BibitemOpen
  \bibfield  {author} {\bibinfo {author} {\bibfnamefont {M.}~\bibnamefont {Modugno}},\ }\href {https://doi.org/10.1088/1367-2630/11/3/033023} {\bibfield  {journal} {\bibinfo  {journal} {New Journal of Physics}\ }\textbf {\bibinfo {volume} {11}},\ \bibinfo {pages} {033023} (\bibinfo {year} {2009})}\BibitemShut {NoStop}%
\bibitem [{\citenamefont {Jitomirskaya}(1999)}]{Jitomirskaya-AAH-Mathematics}%
  \BibitemOpen
  \bibfield  {author} {\bibinfo {author} {\bibfnamefont {S.~Y.}\ \bibnamefont {Jitomirskaya}},\ }\href {http://www.jstor.org/stable/121066} {\bibfield  {journal} {\bibinfo  {journal} {Ann. Math.}\ }\textbf {\bibinfo {volume} {150}},\ \bibinfo {pages} {1159} (\bibinfo {year} {1999})}\BibitemShut {NoStop}%
\bibitem [{\citenamefont {Lucioni}\ \emph {et~al.}(2011)\citenamefont {Lucioni}, \citenamefont {Deissler}, \citenamefont {Tanzi}, \citenamefont {Roati}, \citenamefont {Zaccanti}, \citenamefont {Modugno}, \citenamefont {Larcher}, \citenamefont {Dalfovo}, \citenamefont {Inguscio},\ and\ \citenamefont {Modugno}}]{Lucioni_subdiff_PhysRevLett.106.230403}%
  \BibitemOpen
  \bibfield  {author} {\bibinfo {author} {\bibfnamefont {E.}~\bibnamefont {Lucioni}}, \bibinfo {author} {\bibfnamefont {B.}~\bibnamefont {Deissler}}, \bibinfo {author} {\bibfnamefont {L.}~\bibnamefont {Tanzi}}, \bibinfo {author} {\bibfnamefont {G.}~\bibnamefont {Roati}}, \bibinfo {author} {\bibfnamefont {M.}~\bibnamefont {Zaccanti}}, \bibinfo {author} {\bibfnamefont {M.}~\bibnamefont {Modugno}}, \bibinfo {author} {\bibfnamefont {M.}~\bibnamefont {Larcher}}, \bibinfo {author} {\bibfnamefont {F.}~\bibnamefont {Dalfovo}}, \bibinfo {author} {\bibfnamefont {M.}~\bibnamefont {Inguscio}},\ and\ \bibinfo {author} {\bibfnamefont {G.}~\bibnamefont {Modugno}},\ }\href {https://doi.org/10.1103/PhysRevLett.106.230403} {\bibfield  {journal} {\bibinfo  {journal} {Phys. Rev. Lett.}\ }\textbf {\bibinfo {volume} {106}},\ \bibinfo {pages} {230403} (\bibinfo {year} {2011})}\BibitemShut {NoStop}%
\bibitem [{\citenamefont {Deissler}\ \emph {et~al.}(2010)\citenamefont {Deissler}, \citenamefont {Zaccanti}, \citenamefont {Roati}, \citenamefont {D'Errico}, \citenamefont {Fattori}, \citenamefont {Modugno}, \citenamefont {Modugno},\ and\ \citenamefont {Inguscio}}]{Deissler2010}%
  \BibitemOpen
  \bibfield  {author} {\bibinfo {author} {\bibfnamefont {B.}~\bibnamefont {Deissler}}, \bibinfo {author} {\bibfnamefont {M.}~\bibnamefont {Zaccanti}}, \bibinfo {author} {\bibfnamefont {G.}~\bibnamefont {Roati}}, \bibinfo {author} {\bibfnamefont {C.}~\bibnamefont {D'Errico}}, \bibinfo {author} {\bibfnamefont {M.}~\bibnamefont {Fattori}}, \bibinfo {author} {\bibfnamefont {M.}~\bibnamefont {Modugno}}, \bibinfo {author} {\bibfnamefont {G.}~\bibnamefont {Modugno}},\ and\ \bibinfo {author} {\bibfnamefont {M.}~\bibnamefont {Inguscio}},\ }\href {https://doi.org/10.1038/nphys1635} {\bibfield  {journal} {\bibinfo  {journal} {Nat. Phys.}\ }\textbf {\bibinfo {volume} {6}},\ \bibinfo {pages} {354} (\bibinfo {year} {2010})}\BibitemShut {NoStop}%
\bibitem [{\citenamefont {Eckardt}\ \emph {et~al.}(2005)\citenamefont {Eckardt}, \citenamefont {Weiss},\ and\ \citenamefont {Holthaus}}]{eckardt_superfluid-insulator_2005}%
  \BibitemOpen
  \bibfield  {author} {\bibinfo {author} {\bibfnamefont {A.}~\bibnamefont {Eckardt}}, \bibinfo {author} {\bibfnamefont {C.}~\bibnamefont {Weiss}},\ and\ \bibinfo {author} {\bibfnamefont {M.}~\bibnamefont {Holthaus}},\ }\href {https://doi.org/10.1103/PhysRevLett.95.260404} {\bibfield  {journal} {\bibinfo  {journal} {Phys. Rev. Lett.}\ }\textbf {\bibinfo {volume} {95}},\ \bibinfo {pages} {260404} (\bibinfo {year} {2005})}\BibitemShut {NoStop}%
\bibitem [{\citenamefont {Graham}\ \emph {et~al.}(1992)\citenamefont {Graham}, \citenamefont {Schlautmann},\ and\ \citenamefont {Zoller}}]{Zoller_DL_92}%
  \BibitemOpen
  \bibfield  {author} {\bibinfo {author} {\bibfnamefont {R.}~\bibnamefont {Graham}}, \bibinfo {author} {\bibfnamefont {M.}~\bibnamefont {Schlautmann}},\ and\ \bibinfo {author} {\bibfnamefont {P.}~\bibnamefont {Zoller}},\ }\href {https://doi.org/10.1103/PhysRevA.45.R19} {\bibfield  {journal} {\bibinfo  {journal} {Phys. Rev. A}\ }\textbf {\bibinfo {volume} {45}},\ \bibinfo {pages} {R19} (\bibinfo {year} {1992})}\BibitemShut {NoStop}%
\bibitem [{\citenamefont {Madison}\ \emph {et~al.}(1998)\citenamefont {Madison}, \citenamefont {Fischer}, \citenamefont {Diener}, \citenamefont {Niu},\ and\ \citenamefont {Raizen}}]{Raizen_DL_98}%
  \BibitemOpen
  \bibfield  {author} {\bibinfo {author} {\bibfnamefont {K.~W.}\ \bibnamefont {Madison}}, \bibinfo {author} {\bibfnamefont {M.~C.}\ \bibnamefont {Fischer}}, \bibinfo {author} {\bibfnamefont {R.~B.}\ \bibnamefont {Diener}}, \bibinfo {author} {\bibfnamefont {Q.}~\bibnamefont {Niu}},\ and\ \bibinfo {author} {\bibfnamefont {M.~G.}\ \bibnamefont {Raizen}},\ }\href {https://doi.org/10.1103/PhysRevLett.81.5093} {\bibfield  {journal} {\bibinfo  {journal} {Phys. Rev. Lett.}\ }\textbf {\bibinfo {volume} {81}},\ \bibinfo {pages} {5093} (\bibinfo {year} {1998})}\BibitemShut {NoStop}%
\bibitem [{\citenamefont {Lignier}\ \emph {et~al.}(2007)\citenamefont {Lignier}, \citenamefont {Sias}, \citenamefont {Ciampini}, \citenamefont {Singh}, \citenamefont {Zenesini}, \citenamefont {Morsch},\ and\ \citenamefont {Arimondo}}]{arimondo-shakenlattice}%
  \BibitemOpen
  \bibfield  {author} {\bibinfo {author} {\bibfnamefont {H.}~\bibnamefont {Lignier}}, \bibinfo {author} {\bibfnamefont {C.}~\bibnamefont {Sias}}, \bibinfo {author} {\bibfnamefont {D.}~\bibnamefont {Ciampini}}, \bibinfo {author} {\bibfnamefont {Y.}~\bibnamefont {Singh}}, \bibinfo {author} {\bibfnamefont {A.}~\bibnamefont {Zenesini}}, \bibinfo {author} {\bibfnamefont {O.}~\bibnamefont {Morsch}},\ and\ \bibinfo {author} {\bibfnamefont {E.}~\bibnamefont {Arimondo}},\ }\href {https://doi.org/10.1103/PhysRevLett.99.220403} {\bibfield  {journal} {\bibinfo  {journal} {Phys. Rev. Lett.}\ }\textbf {\bibinfo {volume} {99}},\ \bibinfo {pages} {220403} (\bibinfo {year} {2007})}\BibitemShut {NoStop}%
\bibitem [{\citenamefont {Holthaus}(2016)}]{holthaus_floquet_2016}%
  \BibitemOpen
  \bibfield  {author} {\bibinfo {author} {\bibfnamefont {M.}~\bibnamefont {Holthaus}},\ }\href {https://doi.org/10.1088/0953-4075/49/1/013001} {\bibfield  {journal} {\bibinfo  {journal} {J. Phys. B}\ }\textbf {\bibinfo {volume} {49}},\ \bibinfo {pages} {013001} (\bibinfo {year} {2016})}\BibitemShut {NoStop}%
\bibitem [{\citenamefont {Le}\ \emph {et~al.}(2022)\citenamefont {Le}, \citenamefont {Akashi},\ and\ \citenamefont {Tsuneyuki}}]{Floquet_reform_with_avg_E_22}%
  \BibitemOpen
  \bibfield  {author} {\bibinfo {author} {\bibfnamefont {C.~M.}\ \bibnamefont {Le}}, \bibinfo {author} {\bibfnamefont {R.}~\bibnamefont {Akashi}},\ and\ \bibinfo {author} {\bibfnamefont {S.}~\bibnamefont {Tsuneyuki}},\ }\href {https://doi.org/10.1103/PhysRevA.105.052213} {\bibfield  {journal} {\bibinfo  {journal} {Phys. Rev. A}\ }\textbf {\bibinfo {volume} {105}},\ \bibinfo {pages} {052213} (\bibinfo {year} {2022})}\BibitemShut {NoStop}%
\bibitem [{\citenamefont {Tarallo}\ \emph {et~al.}(2012)\citenamefont {Tarallo}, \citenamefont {Alberti}, \citenamefont {Poli}, \citenamefont {Chiofalo}, \citenamefont {Wang},\ and\ \citenamefont {Tino}}]{Tarallo_PhysRevA.86.033615}%
  \BibitemOpen
  \bibfield  {author} {\bibinfo {author} {\bibfnamefont {M.~G.}\ \bibnamefont {Tarallo}}, \bibinfo {author} {\bibfnamefont {A.}~\bibnamefont {Alberti}}, \bibinfo {author} {\bibfnamefont {N.}~\bibnamefont {Poli}}, \bibinfo {author} {\bibfnamefont {M.~L.}\ \bibnamefont {Chiofalo}}, \bibinfo {author} {\bibfnamefont {F.-Y.}\ \bibnamefont {Wang}},\ and\ \bibinfo {author} {\bibfnamefont {G.~M.}\ \bibnamefont {Tino}},\ }\href {https://doi.org/10.1103/PhysRevA.86.033615} {\bibfield  {journal} {\bibinfo  {journal} {Phys. Rev. A}\ }\textbf {\bibinfo {volume} {86}},\ \bibinfo {pages} {033615} (\bibinfo {year} {2012})}\BibitemShut {NoStop}%
\bibitem [{\citenamefont {Miyake}\ \emph {et~al.}(2013)\citenamefont {Miyake}, \citenamefont {Siviloglou}, \citenamefont {Kennedy}, \citenamefont {Burton},\ and\ \citenamefont {Ketterle}}]{MiyakePRL}%
  \BibitemOpen
  \bibfield  {author} {\bibinfo {author} {\bibfnamefont {H.}~\bibnamefont {Miyake}}, \bibinfo {author} {\bibfnamefont {G.~A.}\ \bibnamefont {Siviloglou}}, \bibinfo {author} {\bibfnamefont {C.~J.}\ \bibnamefont {Kennedy}}, \bibinfo {author} {\bibfnamefont {W.~C.}\ \bibnamefont {Burton}},\ and\ \bibinfo {author} {\bibfnamefont {W.}~\bibnamefont {Ketterle}},\ }\href {https://doi.org/10.1103/PhysRevLett.111.185302} {\bibfield  {journal} {\bibinfo  {journal} {Phys. Rev. Lett.}\ }\textbf {\bibinfo {volume} {111}},\ \bibinfo {pages} {185302} (\bibinfo {year} {2013})}\BibitemShut {NoStop}%
\bibitem [{\citenamefont {Bairey}\ \emph {et~al.}(2017)\citenamefont {Bairey}, \citenamefont {Refael},\ and\ \citenamefont {Lindner}}]{Bairey_PhysRevB.96.020201}%
  \BibitemOpen
  \bibfield  {author} {\bibinfo {author} {\bibfnamefont {E.}~\bibnamefont {Bairey}}, \bibinfo {author} {\bibfnamefont {G.}~\bibnamefont {Refael}},\ and\ \bibinfo {author} {\bibfnamefont {N.~H.}\ \bibnamefont {Lindner}},\ }\href {https://doi.org/10.1103/PhysRevB.96.020201} {\bibfield  {journal} {\bibinfo  {journal} {Phys. Rev. B}\ }\textbf {\bibinfo {volume} {96}},\ \bibinfo {pages} {020201(R)} (\bibinfo {year} {2017})}\BibitemShut {NoStop}%
\bibitem [{\citenamefont {Shimasaki}\ \emph {et~al.}(2024{\natexlab{a}})\citenamefont {Shimasaki}, \citenamefont {Prichard}, \citenamefont {Kondakci}, \citenamefont {Pagett}, \citenamefont {Bai}, \citenamefont {Dotti}, \citenamefont {Cao}, \citenamefont {Dardia}, \citenamefont {Lu}, \citenamefont {Grover},\ and\ \citenamefont {Weld}}]{shimasaki_anomalous_2024}%
  \BibitemOpen
  \bibfield  {author} {\bibinfo {author} {\bibfnamefont {T.}~\bibnamefont {Shimasaki}}, \bibinfo {author} {\bibfnamefont {M.}~\bibnamefont {Prichard}}, \bibinfo {author} {\bibfnamefont {H.~E.}\ \bibnamefont {Kondakci}}, \bibinfo {author} {\bibfnamefont {J.~E.}\ \bibnamefont {Pagett}}, \bibinfo {author} {\bibfnamefont {Y.}~\bibnamefont {Bai}}, \bibinfo {author} {\bibfnamefont {P.}~\bibnamefont {Dotti}}, \bibinfo {author} {\bibfnamefont {A.}~\bibnamefont {Cao}}, \bibinfo {author} {\bibfnamefont {A.~R.}\ \bibnamefont {Dardia}}, \bibinfo {author} {\bibfnamefont {T.-C.}\ \bibnamefont {Lu}}, \bibinfo {author} {\bibfnamefont {T.}~\bibnamefont {Grover}},\ and\ \bibinfo {author} {\bibfnamefont {D.~M.}\ \bibnamefont {Weld}},\ }\href {https://doi.org/10.1038/s41567-023-02329-4} {\bibfield  {journal} {\bibinfo  {journal} {Nat. Phys.}\ }\textbf {\bibinfo {volume} {20}},\ \bibinfo {pages} {409} (\bibinfo {year} {2024}{\natexlab{a}})}\BibitemShut {NoStop}%
\bibitem [{\citenamefont {Walters}\ \emph {et~al.}(2013)\citenamefont {Walters}, \citenamefont {Cotugno}, \citenamefont {Johnson}, \citenamefont {Clark},\ and\ \citenamefont {Jaksch}}]{MLGWS}%
  \BibitemOpen
  \bibfield  {author} {\bibinfo {author} {\bibfnamefont {R.}~\bibnamefont {Walters}}, \bibinfo {author} {\bibfnamefont {G.}~\bibnamefont {Cotugno}}, \bibinfo {author} {\bibfnamefont {T.~H.}\ \bibnamefont {Johnson}}, \bibinfo {author} {\bibfnamefont {S.~R.}\ \bibnamefont {Clark}},\ and\ \bibinfo {author} {\bibfnamefont {D.}~\bibnamefont {Jaksch}},\ }\href {https://doi.org/10.1103/PhysRevA.87.043613} {\bibfield  {journal} {\bibinfo  {journal} {Phys. Rev. A}\ }\textbf {\bibinfo {volume} {87}},\ \bibinfo {pages} {043613} (\bibinfo {year} {2013})}\BibitemShut {NoStop}%
\bibitem [{\citenamefont {Lucioni}\ \emph {et~al.}(2013)\citenamefont {Lucioni}, \citenamefont {Tanzi}, \citenamefont {D'Errico}, \citenamefont {Moratti}, \citenamefont {Inguscio},\ and\ \citenamefont {Modugno}}]{LucioniPRE_PhysRevE.87.042922}%
  \BibitemOpen
  \bibfield  {author} {\bibinfo {author} {\bibfnamefont {E.}~\bibnamefont {Lucioni}}, \bibinfo {author} {\bibfnamefont {L.}~\bibnamefont {Tanzi}}, \bibinfo {author} {\bibfnamefont {C.}~\bibnamefont {D'Errico}}, \bibinfo {author} {\bibfnamefont {M.}~\bibnamefont {Moratti}}, \bibinfo {author} {\bibfnamefont {M.}~\bibnamefont {Inguscio}},\ and\ \bibinfo {author} {\bibfnamefont {G.}~\bibnamefont {Modugno}},\ }\href {https://doi.org/10.1103/PhysRevE.87.042922} {\bibfield  {journal} {\bibinfo  {journal} {Phys. Rev. E}\ }\textbf {\bibinfo {volume} {87}},\ \bibinfo {pages} {042922} (\bibinfo {year} {2013})}\BibitemShut {NoStop}%
\bibitem [{\citenamefont {Shimasaki}\ \emph {et~al.}(2024{\natexlab{b}})\citenamefont {Shimasaki}, \citenamefont {Bai}, \citenamefont {Kondakci}, \citenamefont {Dotti}, \citenamefont {Pagett}, \citenamefont {Dardia}, \citenamefont {Prichard}, \citenamefont {Eckardt},\ and\ \citenamefont {Weld}}]{Shimasaki_PhysRevLett.133.083405}%
  \BibitemOpen
  \bibfield  {author} {\bibinfo {author} {\bibfnamefont {T.}~\bibnamefont {Shimasaki}}, \bibinfo {author} {\bibfnamefont {Y.}~\bibnamefont {Bai}}, \bibinfo {author} {\bibfnamefont {H.~E.}\ \bibnamefont {Kondakci}}, \bibinfo {author} {\bibfnamefont {P.}~\bibnamefont {Dotti}}, \bibinfo {author} {\bibfnamefont {J.~E.}\ \bibnamefont {Pagett}}, \bibinfo {author} {\bibfnamefont {A.~R.}\ \bibnamefont {Dardia}}, \bibinfo {author} {\bibfnamefont {M.}~\bibnamefont {Prichard}}, \bibinfo {author} {\bibfnamefont {A.}~\bibnamefont {Eckardt}},\ and\ \bibinfo {author} {\bibfnamefont {D.~M.}\ \bibnamefont {Weld}},\ }\href {https://doi.org/10.1103/PhysRevLett.133.083405} {\bibfield  {journal} {\bibinfo  {journal} {Phys. Rev. Lett.}\ }\textbf {\bibinfo {volume} {133}},\ \bibinfo {pages} {083405} (\bibinfo {year} {2024}{\natexlab{b}})}\BibitemShut {NoStop}%
\bibitem [{\citenamefont {Martinez}\ and\ \citenamefont {Molina}(2006)}]{Martinez_PhysRevB.73.073104}%
  \BibitemOpen
  \bibfield  {author} {\bibinfo {author} {\bibfnamefont {D.~F.}\ \bibnamefont {Martinez}}\ and\ \bibinfo {author} {\bibfnamefont {R.~A.}\ \bibnamefont {Molina}},\ }\href {https://doi.org/10.1103/PhysRevB.73.073104} {\bibfield  {journal} {\bibinfo  {journal} {Phys. Rev. B}\ }\textbf {\bibinfo {volume} {73}},\ \bibinfo {pages} {073104} (\bibinfo {year} {2006})}\BibitemShut {NoStop}%
\bibitem [{\citenamefont {Kramer}\ and\ \citenamefont {MacKinnon}(1993)}]{LocReview_B_Kramer_1993}%
  \BibitemOpen
  \bibfield  {author} {\bibinfo {author} {\bibfnamefont {B.}~\bibnamefont {Kramer}}\ and\ \bibinfo {author} {\bibfnamefont {A.}~\bibnamefont {MacKinnon}},\ }\href {https://doi.org/10.1088/0034-4885/56/12/001} {\bibfield  {journal} {\bibinfo  {journal} {Rep. Prog. Phys.}\ }\textbf {\bibinfo {volume} {56}},\ \bibinfo {pages} {1469} (\bibinfo {year} {1993})}\BibitemShut {NoStop}%
\bibitem [{\citenamefont {Thouless}(1972)}]{D_J_Thouless_1972}%
  \BibitemOpen
  \bibfield  {author} {\bibinfo {author} {\bibfnamefont {D.~J.}\ \bibnamefont {Thouless}},\ }\href {https://doi.org/10.1088/0022-3719/5/1/010} {\bibfield  {journal} {\bibinfo  {journal} {J. Phys. C}\ }\textbf {\bibinfo {volume} {5}},\ \bibinfo {pages} {77} (\bibinfo {year} {1972})}\BibitemShut {NoStop}%
\end{thebibliography}%


\begin{thebibliography}{11}%
\makeatletter
\providecommand \@ifxundefined [1]{%
 \@ifx{#1\undefined}
}%
\providecommand \@ifnum [1]{%
 \ifnum #1\expandafter \@firstoftwo
 \else \expandafter \@secondoftwo
 \fi
}%
\providecommand \@ifx [1]{%
 \ifx #1\expandafter \@firstoftwo
 \else \expandafter \@secondoftwo
 \fi
}%
\providecommand \natexlab [1]{#1}%
\providecommand \enquote  [1]{``#1''}%
\providecommand \bibnamefont  [1]{#1}%
\providecommand \bibfnamefont [1]{#1}%
\providecommand \citenamefont [1]{#1}%
\providecommand \href@noop [0]{\@secondoftwo}%
\providecommand \href [0]{\begingroup \@sanitize@url \@href}%
\providecommand \@href[1]{\@@startlink{#1}\@@href}%
\providecommand \@@href[1]{\endgroup#1\@@endlink}%
\providecommand \@sanitize@url [0]{\catcode `\\12\catcode `\$12\catcode `\&12\catcode `\#12\catcode `\^12\catcode `\_12\catcode `\%12\relax}%
\providecommand \@@startlink[1]{}%
\providecommand \@@endlink[0]{}%
\providecommand \url  [0]{\begingroup\@sanitize@url \@url }%
\providecommand \@url [1]{\endgroup\@href {#1}{\urlprefix }}%
\providecommand \urlprefix  [0]{URL }%
\providecommand \Eprint [0]{\href }%
\providecommand \doibase [0]{https://doi.org/}%
\providecommand \selectlanguage [0]{\@gobble}%
\providecommand \bibinfo  [0]{\@secondoftwo}%
\providecommand \bibfield  [0]{\@secondoftwo}%
\providecommand \translation [1]{[#1]}%
\providecommand \BibitemOpen [0]{}%
\providecommand \bibitemStop [0]{}%
\providecommand \bibitemNoStop [0]{.\EOS\space}%
\providecommand \EOS [0]{\spacefactor3000\relax}%
\providecommand \BibitemShut  [1]{\csname bibitem#1\endcsname}%
\let\auto@bib@innerbib\@empty
%</preamble>
\bibitem [{\citenamefont {Shimasaki}\ \emph {et~al.}(2024{\natexlab{a}})\citenamefont {Shimasaki}, \citenamefont {Prichard}, \citenamefont {Kondakci}, \citenamefont {Pagett}, \citenamefont {Bai}, \citenamefont {Dotti}, \citenamefont {Cao}, \citenamefont {Dardia}, \citenamefont {Lu}, \citenamefont {Grover},\ and\ \citenamefont {Weld}}]{shimasaki_anomalous_2024}%
  \BibitemOpen
  \bibfield  {author} {\bibinfo {author} {\bibfnamefont {T.}~\bibnamefont {Shimasaki}}, \bibinfo {author} {\bibfnamefont {M.}~\bibnamefont {Prichard}}, \bibinfo {author} {\bibfnamefont {H.~E.}\ \bibnamefont {Kondakci}}, \bibinfo {author} {\bibfnamefont {J.~E.}\ \bibnamefont {Pagett}}, \bibinfo {author} {\bibfnamefont {Y.}~\bibnamefont {Bai}}, \bibinfo {author} {\bibfnamefont {P.}~\bibnamefont {Dotti}}, \bibinfo {author} {\bibfnamefont {A.}~\bibnamefont {Cao}}, \bibinfo {author} {\bibfnamefont {A.~R.}\ \bibnamefont {Dardia}}, \bibinfo {author} {\bibfnamefont {T.-C.}\ \bibnamefont {Lu}}, \bibinfo {author} {\bibfnamefont {T.}~\bibnamefont {Grover}},\ and\ \bibinfo {author} {\bibfnamefont {D.~M.}\ \bibnamefont {Weld}},\ }\href {https://doi.org/10.1038/s41567-023-02329-4} {\bibfield  {journal} {\bibinfo  {journal} {Nat. Phys.}\ }\textbf {\bibinfo {volume} {20}},\ \bibinfo {pages} {409} (\bibinfo {year} {2024}{\natexlab{a}})}\BibitemShut {NoStop}%
\bibitem [{\citenamefont {Drese}\ and\ \citenamefont {Holthaus}(1997)}]{Drese+Holthaus_97}%
  \BibitemOpen
  \bibfield  {author} {\bibinfo {author} {\bibfnamefont {K.}~\bibnamefont {Drese}}\ and\ \bibinfo {author} {\bibfnamefont {M.}~\bibnamefont {Holthaus}},\ }\href {https://doi.org/10.1103/PhysRevLett.78.2932} {\bibfield  {journal} {\bibinfo  {journal} {Phys. Rev. Lett.}\ }\textbf {\bibinfo {volume} {78}},\ \bibinfo {pages} {2932} (\bibinfo {year} {1997})}\BibitemShut {NoStop}%
\bibitem [{\citenamefont {Walters}\ \emph {et~al.}(2013)\citenamefont {Walters}, \citenamefont {Cotugno}, \citenamefont {Johnson}, \citenamefont {Clark},\ and\ \citenamefont {Jaksch}}]{MLGWS}%
  \BibitemOpen
  \bibfield  {author} {\bibinfo {author} {\bibfnamefont {R.}~\bibnamefont {Walters}}, \bibinfo {author} {\bibfnamefont {G.}~\bibnamefont {Cotugno}}, \bibinfo {author} {\bibfnamefont {T.~H.}\ \bibnamefont {Johnson}}, \bibinfo {author} {\bibfnamefont {S.~R.}\ \bibnamefont {Clark}},\ and\ \bibinfo {author} {\bibfnamefont {D.}~\bibnamefont {Jaksch}},\ }\href {https://doi.org/10.1103/PhysRevA.87.043613} {\bibfield  {journal} {\bibinfo  {journal} {Phys. Rev. A}\ }\textbf {\bibinfo {volume} {87}},\ \bibinfo {pages} {043613} (\bibinfo {year} {2013})}\BibitemShut {NoStop}%
\bibitem [{\citenamefont {Lucioni}\ \emph {et~al.}(2011)\citenamefont {Lucioni}, \citenamefont {Deissler}, \citenamefont {Tanzi}, \citenamefont {Roati}, \citenamefont {Zaccanti}, \citenamefont {Modugno}, \citenamefont {Larcher}, \citenamefont {Dalfovo}, \citenamefont {Inguscio},\ and\ \citenamefont {Modugno}}]{Lucioni_subdiff_PhysRevLett.106.230403}%
  \BibitemOpen
  \bibfield  {author} {\bibinfo {author} {\bibfnamefont {E.}~\bibnamefont {Lucioni}}, \bibinfo {author} {\bibfnamefont {B.}~\bibnamefont {Deissler}}, \bibinfo {author} {\bibfnamefont {L.}~\bibnamefont {Tanzi}}, \bibinfo {author} {\bibfnamefont {G.}~\bibnamefont {Roati}}, \bibinfo {author} {\bibfnamefont {M.}~\bibnamefont {Zaccanti}}, \bibinfo {author} {\bibfnamefont {M.}~\bibnamefont {Modugno}}, \bibinfo {author} {\bibfnamefont {M.}~\bibnamefont {Larcher}}, \bibinfo {author} {\bibfnamefont {F.}~\bibnamefont {Dalfovo}}, \bibinfo {author} {\bibfnamefont {M.}~\bibnamefont {Inguscio}},\ and\ \bibinfo {author} {\bibfnamefont {G.}~\bibnamefont {Modugno}},\ }\href {https://doi.org/10.1103/PhysRevLett.106.230403} {\bibfield  {journal} {\bibinfo  {journal} {Phys. Rev. Lett.}\ }\textbf {\bibinfo {volume} {106}},\ \bibinfo {pages} {230403} (\bibinfo {year} {2011})}\BibitemShut {NoStop}%
\bibitem [{\citenamefont {Lucioni}\ \emph {et~al.}(2013)\citenamefont {Lucioni}, \citenamefont {Tanzi}, \citenamefont {D'Errico}, \citenamefont {Moratti}, \citenamefont {Inguscio},\ and\ \citenamefont {Modugno}}]{LucioniPRE_PhysRevE.87.042922}%
  \BibitemOpen
  \bibfield  {author} {\bibinfo {author} {\bibfnamefont {E.}~\bibnamefont {Lucioni}}, \bibinfo {author} {\bibfnamefont {L.}~\bibnamefont {Tanzi}}, \bibinfo {author} {\bibfnamefont {C.}~\bibnamefont {D'Errico}}, \bibinfo {author} {\bibfnamefont {M.}~\bibnamefont {Moratti}}, \bibinfo {author} {\bibfnamefont {M.}~\bibnamefont {Inguscio}},\ and\ \bibinfo {author} {\bibfnamefont {G.}~\bibnamefont {Modugno}},\ }\href {https://doi.org/10.1103/PhysRevE.87.042922} {\bibfield  {journal} {\bibinfo  {journal} {Phys. Rev. E}\ }\textbf {\bibinfo {volume} {87}},\ \bibinfo {pages} {042922} (\bibinfo {year} {2013})}\BibitemShut {NoStop}%
\bibitem [{\citenamefont {Shimasaki}\ \emph {et~al.}(2024{\natexlab{b}})\citenamefont {Shimasaki}, \citenamefont {Bai}, \citenamefont {Kondakci}, \citenamefont {Dotti}, \citenamefont {Pagett}, \citenamefont {Dardia}, \citenamefont {Prichard}, \citenamefont {Eckardt},\ and\ \citenamefont {Weld}}]{Shimasaki_PhysRevLett.133.083405}%
  \BibitemOpen
  \bibfield  {author} {\bibinfo {author} {\bibfnamefont {T.}~\bibnamefont {Shimasaki}}, \bibinfo {author} {\bibfnamefont {Y.}~\bibnamefont {Bai}}, \bibinfo {author} {\bibfnamefont {H.~E.}\ \bibnamefont {Kondakci}}, \bibinfo {author} {\bibfnamefont {P.}~\bibnamefont {Dotti}}, \bibinfo {author} {\bibfnamefont {J.~E.}\ \bibnamefont {Pagett}}, \bibinfo {author} {\bibfnamefont {A.~R.}\ \bibnamefont {Dardia}}, \bibinfo {author} {\bibfnamefont {M.}~\bibnamefont {Prichard}}, \bibinfo {author} {\bibfnamefont {A.}~\bibnamefont {Eckardt}},\ and\ \bibinfo {author} {\bibfnamefont {D.~M.}\ \bibnamefont {Weld}},\ }\href {https://doi.org/10.1103/PhysRevLett.133.083405} {\bibfield  {journal} {\bibinfo  {journal} {Phys. Rev. Lett.}\ }\textbf {\bibinfo {volume} {133}},\ \bibinfo {pages} {083405} (\bibinfo {year} {2024}{\natexlab{b}})}\BibitemShut {NoStop}%
\bibitem [{\citenamefont {Deissler}\ \emph {et~al.}(2010)\citenamefont {Deissler}, \citenamefont {Zaccanti}, \citenamefont {Roati}, \citenamefont {D'Errico}, \citenamefont {Fattori}, \citenamefont {Modugno}, \citenamefont {Modugno},\ and\ \citenamefont {Inguscio}}]{Deissler2010}%
  \BibitemOpen
  \bibfield  {author} {\bibinfo {author} {\bibfnamefont {B.}~\bibnamefont {Deissler}}, \bibinfo {author} {\bibfnamefont {M.}~\bibnamefont {Zaccanti}}, \bibinfo {author} {\bibfnamefont {G.}~\bibnamefont {Roati}}, \bibinfo {author} {\bibfnamefont {C.}~\bibnamefont {D'Errico}}, \bibinfo {author} {\bibfnamefont {M.}~\bibnamefont {Fattori}}, \bibinfo {author} {\bibfnamefont {M.}~\bibnamefont {Modugno}}, \bibinfo {author} {\bibfnamefont {G.}~\bibnamefont {Modugno}},\ and\ \bibinfo {author} {\bibfnamefont {M.}~\bibnamefont {Inguscio}},\ }\href {https://doi.org/10.1038/nphys1635} {\bibfield  {journal} {\bibinfo  {journal} {Nat. Phys.}\ }\textbf {\bibinfo {volume} {6}},\ \bibinfo {pages} {354} (\bibinfo {year} {2010})}\BibitemShut {NoStop}%
\bibitem [{\citenamefont {Martinez}\ and\ \citenamefont {Molina}(2006)}]{Martinez_PhysRevB.73.073104}%
  \BibitemOpen
  \bibfield  {author} {\bibinfo {author} {\bibfnamefont {D.~F.}\ \bibnamefont {Martinez}}\ and\ \bibinfo {author} {\bibfnamefont {R.~A.}\ \bibnamefont {Molina}},\ }\href {https://doi.org/10.1103/PhysRevB.73.073104} {\bibfield  {journal} {\bibinfo  {journal} {Phys. Rev. B}\ }\textbf {\bibinfo {volume} {73}},\ \bibinfo {pages} {073104} (\bibinfo {year} {2006})}\BibitemShut {NoStop}%
\bibitem [{\citenamefont {Kramer}\ and\ \citenamefont {MacKinnon}(1993)}]{LocReview_B_Kramer_1993}%
  \BibitemOpen
  \bibfield  {author} {\bibinfo {author} {\bibfnamefont {B.}~\bibnamefont {Kramer}}\ and\ \bibinfo {author} {\bibfnamefont {A.}~\bibnamefont {MacKinnon}},\ }\href {https://doi.org/10.1088/0034-4885/56/12/001} {\bibfield  {journal} {\bibinfo  {journal} {Rep. Prog. Phys.}\ }\textbf {\bibinfo {volume} {56}},\ \bibinfo {pages} {1469} (\bibinfo {year} {1993})}\BibitemShut {NoStop}%
\bibitem [{\citenamefont {Thouless}(1972)}]{D_J_Thouless_1972}%
  \BibitemOpen
  \bibfield  {author} {\bibinfo {author} {\bibfnamefont {D.~J.}\ \bibnamefont {Thouless}},\ }\href {https://doi.org/10.1088/0022-3719/5/1/010} {\bibfield  {journal} {\bibinfo  {journal} {J. Phys. C}\ }\textbf {\bibinfo {volume} {5}},\ \bibinfo {pages} {77} (\bibinfo {year} {1972})}\BibitemShut {NoStop}%
\bibitem [{\citenamefont {Aubry}\ and\ \citenamefont {Andr{\'e}}(1980)}]{aubry1980analyticity}%
  \BibitemOpen
  \bibfield  {author} {\bibinfo {author} {\bibfnamefont {S.}~\bibnamefont {Aubry}}\ and\ \bibinfo {author} {\bibfnamefont {G.}~\bibnamefont {Andr{\'e}}},\ }\href@noop {} {\bibfield  {journal} {\bibinfo  {journal} {Ann. Israel Phys. Soc.}\ }\textbf {\bibinfo {volume} {3}},\ \bibinfo {pages} {18} (\bibinfo {year} {1980})}\BibitemShut {NoStop}%
\end{thebibliography}%

\end{document}

\preprint{APS/123-QED}

\title{Supplementary Material:\\Localization phase diagram controlled by the interplay of disorder and driving}

\author{Peter Dotti}
 % \email{p\_dotti@ucsb.edu}
\author{Yifei Bai}%
\author{Toshihiko Shimasaki}
\author{Anna R. Dardia}
\author{David Weld}%
% \email{weld@ucsb.edu}
\affiliation{%
\textbf{}
Department of Physics, University of California, Santa Barbara, California 93106, USA
}%

% \collaboration{MUSO Collaboration}%\noaffiliation

%  \homepage{http://www.Second.institution.edu/~Charlie.Author}
% \affiliation{
%  Second institution and/or address\\
%  This line break forced% with \\
% }%
% \affiliation{
%  Third institution, the second for Charlie Author
% \affiliation{%
%  Authors' institution and/or address\\
%  This line break forced with \textbackslash\textbackslash
% }%

% \collaboration{CLEO Collaboration}%\noaffiliation

\date{\today}% It is always \today, today,

\maketitle

%Numbering figures with S
\renewcommand\thefigure{S\arabic{figure}} 

% Revtex command to switch to single column
\onecolumngrid
\section{Experimental Details and Tight-Binding Model Parameters}
The experiments described in the main text begin by loading a Bose-Einstein condensate of $^{84}$Sr atoms, numbering $8\times 10^4$ atoms, into a pair of superimposed coaxial 1D optical lattices with different spatial periods, aligned along an axis perpendicular to gravity.  
The deeper of these two lattices is referred to as the primary lattice.  It is generated by counterpropagating Gaussian laser beams with wavelength $\lambda_p=1064$~nm. The two primary lattice beams are derived from the same fiber amplifier, but their frequency and amplitude can be tuned independently using acousto-optic modulators (AOMs).  One of the primary lattice beams has a much higher intensity than any of the other beams (approximately 2~W) in order to support the atoms against gravity.  The depth of the primary lattice $V_p$ is varied using the intensity and polarization of the weaker beam. The polarization was set once initially and kept constant in all stages of the experiment.  $V_p$ was controlled during cooling and loading into the lattice with feedback to power monitoring photodiodes for each of the beams.  

All beams forming the lattices are focused to the point where the BEC is initially formed and loaded into the lattice.  {Because the beams are focused to a narrow beam width at the atoms, the beam intensities vary in space and give rise to a confining potential.  We model this confinement} as a harmonic potential with axial trapping frequency $\omega_r$ in directions transverse to the direction of propagation, and a longitudinal trapping frequency $\omega_z$ along the direction that the beam propagates.   We estimate these trapping frequencies to be $\omega_r = 2\pi\times 100.1 \, \mathrm{Hz}$ and $\omega_z = 2\pi\times 0.307 \, \mathrm{Hz}$, based on measurements of the strong primary lattice beam intensity and its Gaussian beam profile.  These conditions of relatively weak transverse confinement result in atom cloud densities that are sufficiently low that we can describe most of our results in reference to non-interacting models with some noted deviation as in the observed sub-diffusion of transport measurements. {We note that $1/\omega_r$ is sufficiently large compared to the time over which the lattice potential is applied that the initial state is well described by a product state between the 2D ground state of the transverse confinement and the 1D state along the lattice.  Furthermore, the dynamics do not significantly disturb the transverse component of the ground state.  These facts enable us to model the system with a 1D Hamiltonian for the 1D state along the lattice axis in the product state.} {In previous work \cite{shimasaki_anomalous_2024} we investigated how varying the atom density by changing the atom number affects transport behavior in a kicked quasicrystal and found the general structure of the localization phase diagrams reported therein to be qualitatively unchanged, 
the main effect being increased cloud widths with higher densities. We conclude that interparticle interactions do not give rise to the transport features we observe in our quasicrystal experiments.}
All experiments in this work were conducted with $V_p=9E_{R,p}$, where $E_{R,p} = \hbar^2k_p^2/2m$ and $k_p = 2\pi/\lambda_p$.  The depth of the primary lattice was calibrated to be accurate within 5\% of the nominal value of $9E_{R,p}$ by checking the resonant transition frequency between the ground state of the lowest band and the highest energy state of the first excited band for atoms loaded into the lattice using a sinusoidal translation of the lattice at several frequencies near resonance and observing reduction in atom density that results from resonant excitation of atoms from the ground band to the first excited band.  The amplitude of the drive corresponded to $\Delta\nu_{\mathrm{max},p} = 40~\mathrm{Hz}$ (see below.)   The secondary lattice depth was calibrated to an accuracy of 5\% using Kapitza-Dirac diffraction of the atom cloud initially held in the strong beam of the primary lattice.  

The two laser beams that form the secondary lattice have wavelength {$\lambda_s=874.61$~nm} and are derived from a continuous wave titanium sapphire laser stabilized against drift with a wavemeter.  The secondary lattice beams are of roughly equal intensity, are stabilized with monitoring photodiodes, and are adjusted together to control the depth of the secondary lattice $V_s$. As was the case for the primary lattice beams, the secondary lattice beam frequencies are independently tunable by AOMs.  Values of the secondary lattice depth used in the presented experiments range from 0$E_{R,p}$ to 0.59$E_{R,p}$.

This setup provides fast tunability of the potential depths and translation of the optical lattice potentials along the lattice direction, which enables the simultaneous experimental realization of dynamic and quasidisorder-induced localization. The system model is depicted schematically in Fig.~1(a) of the main text and described by the one-dimensional Hamiltonian
\begin{equation}
H_\mathrm{Exp} = -\frac{\hbar^2}{2m}\frac{\mathrm{d}^2}{\mathrm{d}x^2} + V_p\sin^2(k_p (x - \delta_p)) + V_s\sin^2(k_s (x - \delta_s)),
\label{basicHamiltonian}
\end{equation}
where $V_p$ and $V_s$ are the depths of the primary and secondary lattice, respectively. $k_{p(s)} = 2\pi/\lambda_{p(s)}$ and $\delta_{p(s)}$ denotes the possibly time-dependent spatial offset of the of the primary (secondary) lattice, which is determined by the phase difference $\Delta\varphi_{p(s)}$ between the counterpropagating laser beams according to the formula
\begin{equation}
    \delta_{p(s)} = \frac{\Delta\varphi_{p(s)}}{2k_{p(s)}}.
    \label{eqn:latticePosition}
\end{equation}
The velocity of each lattice is thus
\begin{equation}
    v_{p(s)} = \frac{\mathrm{d}}{\mathrm{d}t}\delta_{p(s)} = \frac{\lambda_{p(s)}}{2}\Delta\nu_{p(s)},
    \label{eqn:latticeVelocity}
\end{equation}
where $\Delta\nu_{p(s)}$ is the frequency difference between the two laser beams forming the primary (secondary) lattice.  We directly control $\Delta\nu_{p(s)}$ by adjusting the phase of the radio frequency (RF) drive to one of the AOMs.  All RF waveform generators share a clock signal so that the relative position of the two lattices is well controlled up to unavoidable thermal drift of optical path lengths due to opto-mechanical element drift and index of refraction changes that are unimportant on the time scale of the collection of individual data points, but can change the relative positions of the lattices between runs.  This effectively results in averaging over system realizations with different initial positions ($\delta_{p}$ and $\delta_{s}$ at the initial time) of the lattice in the experimental data.  The time evolution behavior is expected generically to be insensitive to this relative lattice position change, and the data does not measurably change as a result of this drift.

%%%%%%%%%%%%%%%%%%%%%%%%%%%%%%%%%%%%%%%%%%%%%%%%%%%%
%%%%% Tight-Binding Model Description %%%%%
We specialize this paper to synchronized sinusoidal driving of the primary and secondary lattices, so that 
\begin{equation}
    \delta_p = \delta_s + \delta_0 = -A\cos(\omega t) 
\end{equation}
where $\delta_0$ is constant over the time scale of each experimental run, $\omega$ is the angular drive frequency, and $A$ is the drive amplitude.  As stated in the text, we use as our model Hamiltonian the tight-binding model approximation corresponding to Hamiltonian \eqref{basicHamiltonian}, given as~\cite{Drese+Holthaus_97}
\begin{equation}
\begin{split}
H =&\ \sum_{l} -J \Big[\, \ket{l+1}\!\bra{l}\,+\, \ket{l}\!\bra{l+1}\,\Big] + \Delta \sum_{l}\cos(2\pi\beta l-\delta)\ket{l}\!\bra{l} + \hbar \omega K_0 \cos(\omega t)\sum_{l}l\ket{l}\!\bra{l},
\label{basicHamiltonian_TB}
\end{split}
\end{equation}
where $J$ is the tunneling energy, $\Delta$ is the quasidisorder strength, $K_0$ is the dimensionless drive strength, $\beta = \lambda_p/\lambda_s=1064/874.61$ and $\delta=2\delta_0/\lambda_p$. This Hamiltonian describes the system in the reference frame that is comoving with the lattices, i.e.\!\ the lattices are stationary in this frame.  The Hamiltonian in the comoving frame is related to the lab frame by a unitary transformation, consideration of which produces the following relations between $K_0$ and drive amplitude.
\begin{equation}
     K_0 \frac{\hbar\omega}{(\lambda_p/2)}= m A \omega^2 = \frac{m\lambda_p\omega}{2} \Delta\nu_{\mathrm{max},p}.
\label{fictitiousForceTerm}
\end{equation}
Here, $\Delta\nu_{\mathrm{max},p}$ is the experimentally relevant amplitude in frequency difference between the the laser beams forming the primary lattice, i.e.\!\ $\Delta\nu_p = \Delta\nu_{\mathrm{max},p}\sin(\omega t)$.  Each ground band Wannier state $\ket{l}$ corresponds to the state localized at the position $x_l = l\pi {/} k_p + \delta_p$. Wannier states are computed numerically to determine $J$ and $\Delta$~\cite{MLGWS}.  For $V_p=9E_{R,p}$, we have $J=0.024E_{R,p}$, and for the range of $V_s$ values, $\Delta$ ranges from $0E_{R,p}$ to $0.16E_{R,p}$.

Localization properties are observed experimentally via measured width, using the standard deviation $\sigma$ of a best-fit Gaussian to the 1D density distribution, computed from the 2D imaged density distribution by integrating along the direction transverse to the lattice axis.    
In the measured phase diagrams in the main text, the average fitted $\sigma$ of three repeats is shown for each pair of $K$ and $\Delta/J$ values, with the exception of rare (0.1\%) outlier cases in which atom loss resulted in a poor fit, in which cases the remaining two repeats were averaged.  {The initial width of the cloud was measured to be $\sigma_0 = 10\pm 1$~\textmu m.}

{The atomic distribution is measured using absorption imaging through an imaging system with a magnification of 8.  The imaging optics are diffraction-limited with a minimum resolvable distance of 9~\textmu m (approximately 4 camera pixels.)  We are able to accurately measure cloud widths of $\sigma \ge 10$~\textmu m (full width half max $\ge$ 24~\textmu m). Any aberrations in the imaging system are static and are thus not expected to affect the measurement of the change in atomic cloud width.} 

% added 
\subsection{Effect of interaction}
{${}^{84}$Sr-${}^{84}$Sr interaction has a scattering length of $a_s = 123a_0$ where $a_0$ is the Bohr radius. Because of the relatively weak transverse harmonic confinement, the condensate will have an overall broad transverse profile and thus the effect of interaction can be modeled by a nonlinear, mean-field interaction term. For the initial cloud, the transverse width is fitted to be $6.5$~\textmu m. The initial peak density can thus be estimated to be $12\times 10^{12} \, \mathrm{cm}^{-3}$. The dynamics thus can be modeled by the nonlinear Schr\"odinger equation}
$
i\hbar \partial_t \psi_j = -J\left( \psi_{j+1} + \psi_{j-1} \right) + V(t) \psi_j + g \vert 
\psi_j \vert^2\psi_j,
$
{where $V(t)$ describes the time-dependent potential energy including the quasiperiodic potential and the oscillating force, and $g$ is the mean-field interaction strength.} 

{Following \cite{Lucioni_subdiff_PhysRevLett.106.230403, LucioniPRE_PhysRevE.87.042922}, one can estimate the interaction strength by $ E_\mathrm{int} = U \bar{N}/2 = 2Jg/n_s$ where $\bar N$ is the average number of atom per site and $n_s$ is the average number of occupied sites. Thus we can estimate $g = UN/4$, where $N=\bar{N}n_s$ is the total atom number and the Bose-Hubbard interaction parameter $U$ is estimated by $U = (4\pi\hbar^2 a_s/m)\int \vert \psi(\mathbf{r})\vert^4 d^3 \mathbf{r}$, where $m$ is the mass of the ${}^{84}$Sr atom and $\psi(\mathbf{r})$ is the single-site wave function of the atom. We take $\psi(\mathbf{r})$ to be the Gaussian approximation assuming no interactions. This estimate gives $U \approx 0.00188J$ and $g \approx 37.58J$ for our initial condition.}

{For a strong quasiperiodic potential that produces a localized phase in the non-interacting limit, the addition of the mean-field interaction leads to sub-diffusive wave packet expansion \cite{Lucioni_subdiff_PhysRevLett.106.230403}. However, the speed of expansion is sharply reduced compared to the ballistic case where the quasiperiodic potential is weak, as has been observed both numerically and experimentally \cite{Shimasaki_PhysRevLett.133.083405, shimasaki_anomalous_2024, Deissler2010}. In particular, we have observed before \cite{shimasaki_anomalous_2024} that the basic structure of the phase diagram is qualitatively unchanged over a wide range of interaction parameter $g$, both in numerically and experimentally. This is partly explained by our experimental setup: as the interaction increases, the transverse size of the BEC also increases because the transverse confinement is relatively weak, and therefore reduces the overall density. Thus overall, despite the existence of interaction, wave packet expansion still serves a meaningful metric to probe the underlying phase in the underlying metric.} 

\section{Calculation and scaling of the inverse participation ratio}
To determine the Floquet eigenstates and their inverse participation ratios, let us first note that for a generic time-periodic Hamiltonian $H(t) = H(t+T)$ with period $T$, there exists a complete and orthornormal basis $\{ \vert \psi_n (t) \rangle \}$. We refer to the elements of this basis set $\vert \psi_n (t) \rangle$ as Floquet eigenstates. In analogy to the Bloch state in a spatially periodic system, the Floquet eigenstate can be decomposed into a product of a plane wave $e^{-i \varepsilon_n t}$ and a time-periodic ``Floquet function" $\vert \Phi_n (t) \rangle$ as $\vert \psi_n (t) \rangle = e^{-i \varepsilon_n t} \vert \Phi_n (t) \rangle$, where $\varepsilon_n$ is the quasienergy of the Floquet eigenstate. 

The Floquet function has the same time-periodicity of the Hamiltonian: $\vert \Phi_n (t) \rangle = \vert \Phi_n (t+T) \rangle$. We can thus apply the discrete Fourier transform to $\vert \Phi_n (t) \rangle$ and $H(t)$ as
\[
H(t) = \sum_{m=-\infty}^\infty e^{-im\omega t} H_m, \quad \vert \Phi_n (t) \rangle = \sum_{m=-\infty}^\infty e^{-im\omega t} \vert\phi_{m,n}\rangle. 
\]
where $\omega = 2\pi/T$. Once we substitute the discrete Fourier transform into the time-dependent Schr\"odinger equation, we arrive at a time-independent equation for each Fourier component as
\begin{equation}
    (\varepsilon_n + m \hbar\omega)\vert\phi_{m,n}\rangle = \sum_{m'=-\infty}^\infty H_{m-m'}\vert\phi_{m',n}\rangle. \label{eq:Flq-time-ind}    
\end{equation}
We can then obtain quasienergies and Floquet eigenstates by numerically diagonalizing Eq.~(\ref{eq:Flq-time-ind}) followed by the inverse discrete Fourier transform at time $t$.  In analogy to the Brillouin zone in the extended scheme of a crystal, the solution of Eq.~(\ref{eq:Flq-time-ind}) has infinitely-many identical copies of the same Floquet eigenstates, leading to the concept of Floquet-Brillouin zones. We consider the Floquet eigenstates and quasienergy in the central Floquet-Brillouin zone $(m=0)$ for the main text and remainder of this discussion. Finally, in practice, we can only include finitely many Floquet-Brillouin zones in our numerical study.  Hence, the summation in Eq.~(\ref{eq:Flq-time-ind}) is restricted to a finite number of terms with $|m'|\le m_\mathrm{cutoff}$. 

We will now discuss our specific Hamiltonian of interest (Eq. (1) in the main text). We first apply a unitary gauge transformation to the driven AAH Hamiltonian as: 
\[
H \rightarrow H' = -J \sum_{l}\Big[\, e^{i K_0 \sin(\omega t)} \ket{l+1}\!\bra{l}\,+\, e^{-i K_0 \sin(\omega t)} \ket{l}\!\bra{l+1}\,\Big] + \Delta \sum_{l}\cos(2\pi\beta l-\delta)\ket{l}\!\bra{l}. 
\]
The Fourier components of the transformed Hamiltonian $H' = \sum_m e^{-im\omega t} H_m$ is given by the Jacobi-Anger expansion
\[
H_{m} = -J\sum_{l}\Big[\, \mathcal{J}_m (K_0) \ket{l+1}\!\bra{l}\,+\, \mathcal{J}_{-m} (K_0) \ket{l}\!\bra{l+1}\,\Big] + \begin{cases}
    \Delta \sum_{l}\cos(2\pi\beta l-\delta)\ket{l}\!\bra{l}, & m = 0, \\
    0, & m \neq 0.
\end{cases}
\]
The $m=0$ term corresponds to the time-averaged Hamiltonian. We can then substitute $H_m$ into Eq.~(\ref{eq:Flq-time-ind}) and obtain the quasienergies and Floquet eigenstates.

We use the inverse participation ratio (IPR) in the position basis to probe the localization property of the Floquet eigenstates. After projecting the $t=0$ Floquet eigenstate $\vert\psi_i\rangle\equiv \vert\psi_i(t=0)\rangle$ onto the position basis as $\vert\psi_i\rangle = \sum_l \psi_l^{(i)} \vert l \rangle$, its IPR for the $i^\mathrm{th}$ eigenstate is then defined as
\begin{equation}
    \mathrm{IPR}^{(i)} = \frac{\sum_{l=1}^L \vert \psi_l^{(i)} \vert^4}{ \left(\sum_{l=1}^L \vert \psi_l ^{(i)} \vert^2 \right)^2}.
\end{equation}
Here $L$ is the system size. A localized (delocalized) wavefunction has a non-vanishing (vanishing) value of IPR. 

The average IPR, denoted $\langle \mathrm{IPR} \rangle$ in the text, is then defined as
\[
\langle \mathrm{IPR} \rangle = \frac{1}{L} \sum_{i=1}^L \mathrm{IPR}^{(i)}
\]

Furthermore, scaling the IPR with respect to the system size $L$ allows us to distinguish the phases without relying on the numerical values of IPR. The IPR scales as 
\[
\mathrm{IPR}^{(i)} \sim L^{-D_2}.
\]
$D_2$ is called the fractal dimension. When $D_2 = 0 (1)$, the state is localized (delocalized). When $0<D_2<1$, the state is critical with multifractal characteristics. 

\begin{figure}
    \centering
    \includegraphics[width=3.375in]{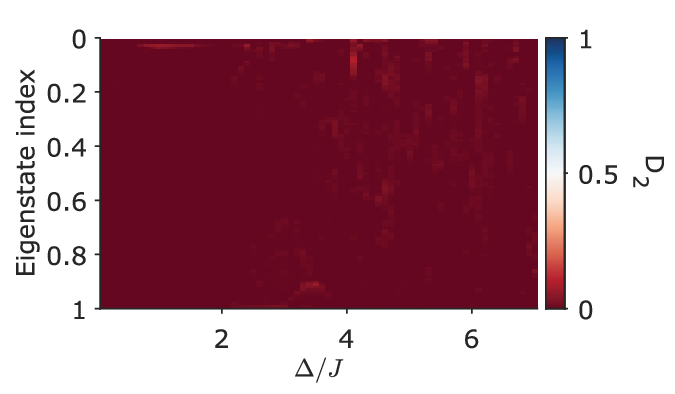}
    \caption{Fractal dimensions of Floquet eigenstates at $K_0 = 2.4046$. }
    \label{fig:D2}
\end{figure}

We computed $D_2$ at $K_0 = 2.4046$ and $\omega = 2\pi\times 200\, \mathrm{Hz}$. The result in Fig.~\ref{fig:D2} shows that all the Floquet eigenstates are localized without any mobility edges or critical states. Thus the observed low-frequency deviation is not explained as a consequence of the anomalous localization observed in \cite{shimasaki_anomalous_2024}.

\section{Energy-dependent localization length with the Floquet-Green formalism}
Here we include numerical calculation of the localization length of the driven AAH model. We vary the quasiperiodic strength $\Delta/J$ and fix $K_0 \approx 2.4046$, the same as the experimental condition for Fig. 4 in the main text.

The localization length is computed via the Floquet-Green formalism \cite{Martinez_PhysRevB.73.073104}. We first recall some basic definitions and background before discussing the actual calculation. We start with a 1D, time-independent lattice system of length $L$ and a particle with energy $E$. The {system's behavior} can be captured by its Green's function $G(E)$. The localization length of the system $\xi(E)$ can then be defined as \cite{LocReview_B_Kramer_1993}
\[
\frac{1}{\xi(E)} = -\lim_{L \rightarrow \infty} \frac{1}{L} \Big\langle\!\ln |{G_{1,L} (E)}| \Big\rangle,
\]
where $G_{1,L} (E)$ is the matrix element of the Green's function at $(1,L)$. Here $\langle \cdot \rangle$ represents averaging over different realizations of the potential. One can relate this quantity to the transmission coefficient of a particle with incident energy $E$ at site $1$ to the end at the site $L$. From this interpretation, $E$ need not be the eigenenergy of the system. From this definition, it is then straightforward to show that the localization length of the static AAH model is $\xi^{-1} = \ln (\Delta/(2J))$ \cite{D_J_Thouless_1972, aubry1980analyticity}.

The generalization of the Green's function to a time-periodic system was discussed in \cite{Martinez_PhysRevB.73.073104}. We consider the Hamiltonian
\[
\hat H(t) = \hat H_\mathrm{static} + 2 \hat V \cos \omega t.
\]
The factor of $2$ is for convenience only. For our model under study, $\hat H_\mathrm{static}$ is the static AAH model, and $\hat V := \hbar\omega K_0/2\sum_l l|l\rangle \langle l |$ represents the oscillating dipolar force. 

The Floquet-Green's function can then be defined with the Fourier components of the Floquet eigenstates
\[
G^{(k)}(E) = \sum_{n,m} \frac{\ket{\phi_{m+k, n}}\bra{\phi_{m,n}}}{E - \epsilon_n - m \hbar \omega}
\]
Similarly to the undriven case, the matrix element $G^{(k)}_{1,L}(E)$ of the Floquet-Green's function relates to the transmission of a particle with energy $E$ through the driven lattice system: the particle enters the lattice with energy $E$ and leaves with energy $E +k\hbar\omega$. The integer $k$ thus labels such a transmission channel. 

Thus, we can extend the definition of the localization length as
\begin{equation}
    \frac{1}{\xi ^{(k)}(E)} = -\lim_{L \rightarrow \infty} \frac{1}{L} \left\langle \ln \left\vert {G_{1,L}^{(k)} (E)} \right\vert \right\rangle.
\end{equation}
It can be shown that $\xi^{(k)}(E)$ are all equal for different integer $k$. Furthermore, in the limit of vanishing driving $\hat V \rightarrow 0$, $G^{(k=0)}(E) \rightarrow G(E)$ and $G^{(k\neq0)}(E) \rightarrow 0$~\cite{Martinez_PhysRevB.73.073104}. Therefore, we can restrict the calculation to ${G^{(0)} (E)}$ and consider $\xi^{(0)}$ only. 

The Floquet-Green's function ${G^{(0)} (E)}$ satisfies \cite{Martinez_PhysRevB.73.073104}
\begin{equation}
    G^{(0)} (E) = \frac{1}{E \hat I - \hat H_{\mathrm{static}} - \hat V_\mathrm{eff} (E)}, \quad \hat V_\mathrm{eff} (E) = \hat V_\mathrm{eff}^+ (E) + \hat V_\mathrm{eff}^- (E),
\end{equation}
where the dynamical effective potential can be evaluated by a matrix continued fraction
\begin{equation}
    \hat V_\mathrm{eff}^\pm (E) = \hat V \frac{1}{ (E \pm 1\hbar \omega)\hat I - \hat H_\mathrm{static} - \hat V \frac{1}{(E \pm 2\hbar \omega)\hat I - \hat H_\mathrm{static} - \hat V \frac{1}{ \vdots}\hat V}\hat V}\hat V.
\end{equation}
In practice, we choose $L = 100$ and iterate the continued fraction 200 times to ensure convergence. We choose $\delta = 0$ in our calculations. 

\begin{figure}[htbp]
    \centering
    \includegraphics[width=3.375in]{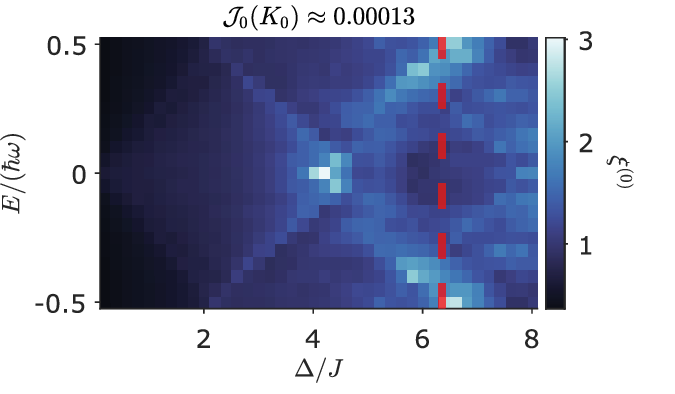}% Here is how to import EPS art
    \caption{Localization length at $\omega = 2 \pi \times 200\, \mathrm{Hz}$ and $K_0 = 2.4046$ in units of the lattice spacing. The dashed red line shows the quasiperiodic strength at which the experiments shown in Fig. 4 in the main text were conducted. }
    \label{fig:locLength}
\end{figure}

We present the result for $\omega = 2 \pi \times 200\, \mathrm{Hz}$ in Fig.~\ref{fig:locLength}. At $K_0 \approx 2.4046$, we expect the localization length to essentially vanish within the high-frequency approximation. This is because the effective quasiperiodic strength $\Delta/(J\mathcal{J}_0(K_0))$ diverges as $K_0$ approaches the zeroes of $\mathcal{J}_0(K_0)$. We observe that the localization lengths are energy dependent at $\omega = 2 \pi \times 200\, \mathrm{Hz}$ and are larger than the high-frequency prediction. This corroborates the results of the IPR calculation. 

\section{Extended discussion of second anomalous feature}
%pasting most of entire discussion here for now, will cut down and tie together more nicely
%transport exponent details
The transport exponents $\gamma$ for each drive frequency from the data in Fig. 4 {of the main text} are extracted by fitting to the form $\sigma(t) = \sigma_0 (1+t/t_0)^\gamma$ \cite{Lucioni_subdiff_PhysRevLett.106.230403}.
%The results are given in Table~\ref{table:transportExp}.   
Here $\sigma_0$ is the initial width, and $t_0$ acts as an activation time. Localization, diffusion, and ballistic expansion correspond to $\gamma = 0$, $0.5$, and $1$, respectively.  Expansion for the static quasiperiodic case exhibits slow sub-diffusion with a transport exponent $\gamma = 0.24(5)$, deviating from perfect localization due to finite mean-field interaction~\cite{Lucioni_subdiff_PhysRevLett.106.230403,shimasaki_anomalous_2024}.  Expansion remains sub-diffusive when the drive is applied and we observe the expansion exponent to be further reduced at high frequencies.  Note that we do not fit the periodic case because {the weak trapping potential with trapping frequency $\omega_z = 2\pi\times 0.307 \, \mathrm{Hz}$ along the lattice} complicates the dynamics at long evolution time.  These experimental results imply that the anomalous region in the lower-right of the phase diagram is still a localized phase. 
%numerical things
We corroborate the experimental results with numerical studies of Eq.~\ref{basicHamiltonian_TB}, for which we again fix $K_0 \approx 2.4046$, but vary the quasidisorder strength $\Delta$. We numerically diagonalize the time-dependent Hamiltonian in Eq.~\ref{basicHamiltonian_TB} via the Floquet formalism and calculate the IPR of the eigenstates in the central Floquet-Brillouin zone. By calculating the scaling of the IPR versus system size, we can determine the phase to be fully localized without mobility edges. 

\begin{figure}[th!]
\includegraphics[width=3.375in]{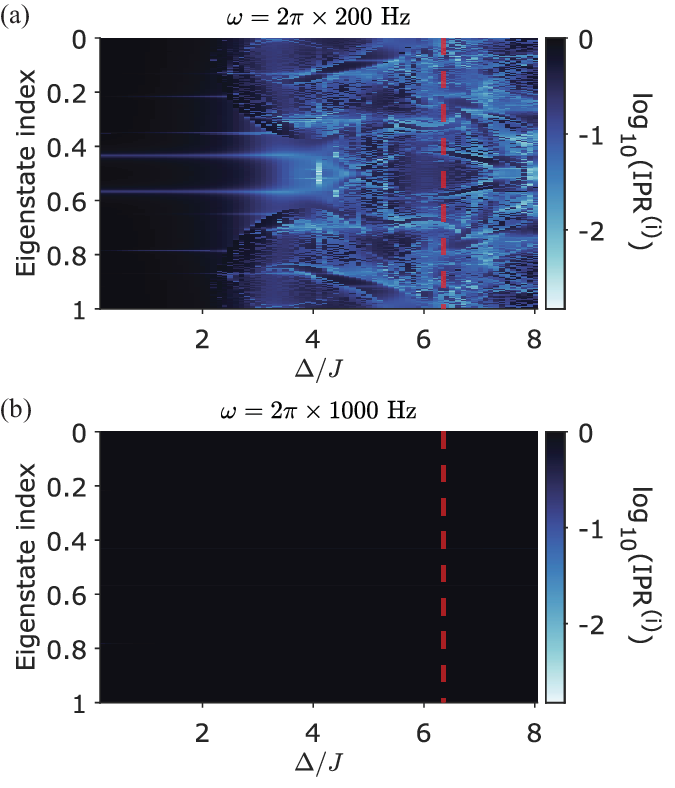}% Here is how to import EPS art
\caption{\label{fig:IPR_energydependent} IPR of Floquet eigenstates in the central Floquet-Brillouin zone in (a) low and (b) high-frequency regimes. Values of IPR of each eigenstate fluctuate heavily for the low-frequency case in (a), but they are independent of each eigenstate in the high-frequency case in (b) up to numerical precision and finite-size effects, {which leads to the completely uniform appearance}. The red dashed lines correspond to the quasiperiodic strength $\Delta/J$ used in the experiments of Fig. 4 in the main text.}
\end{figure}

%IPR
What then distinguishes the anomalous region observed in experiment? We argue that this region features localization lengths which are both quantitatively larger and much more quasienergy dependent than in the high-frequency phase diagram.  Numerically calculated values of IPR in the low-frequency case (Fig.~\ref{fig:IPR_energydependent}(a)) are smaller than those in the high-frequency case (Fig.~\ref{fig:IPR_energydependent}(b)), meaning that the localization lengths in the low-frequency regime are larger. Previous numerical studies showed that localization lengths increase under low-frequency driving for a tight-binding lattice with truly random disorder~\cite{Martinez_PhysRevB.73.073104}. Our calculations thus generalize this result to a quasiperiodic system in the localized phase. The IPR of the low-frequency Floquet eigenstates also vary with quasienergy, as shown in Fig.~\ref{fig:IPR_energydependent}(a) and discussed from a different perspective in the preceding section.%~\cite{supp_mat}. 

\section{300 Hz Drive Frequency Data Used in Fig.~3(\lowercase{b})}

\begin{figure}[htbp]
    \centering
    \includegraphics[width=3.375in]{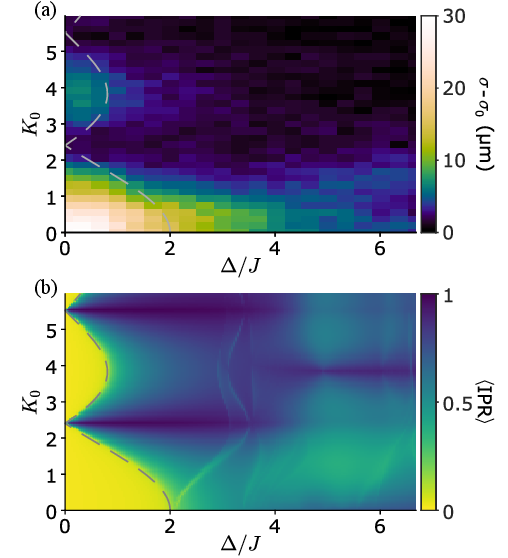}% Here is how to import EPS art
    \caption{Experimental and numerical simulation results for the system driven at frequency $\omega = 2\pi\times300\mathrm{\ Hz}$, presented as a function of system parameters $K_0$ and $\Delta/J$. (a)~Measurements of {$\sigma-\sigma_0$}.  (b) Calculated $\langle \mathrm{IPR}\rangle$.}
    \label{fig:300HzPhaseDiagrams}
\end{figure}

In Fig.~\ref{fig:300HzPhaseDiagrams} we present data for the measured {change in width $\sigma-\sigma_0$ alongside the} calculated $\langle \mathrm{IPR}\rangle$ for the system driven at the frequency $\omega = 2\pi\times300\mathrm{\ Hz}$.  The data shown is analogous to that presented in Fig.~2 of the main text and was obtained using the same procedures.  The data shown in Fig.~\ref{fig:300HzPhaseDiagrams}(a) was used in conjunction with the $\omega = 2\pi\times1\mathrm{\ kHz}$ data shown in Fig.~2(a) of the main text to calculate the ratios, denoted $\sigma_\text{300Hz}/\sigma_\text{1kHz}$, that are presented in Fig.~3(b) of the main text.

%\clearpage

\bibliography{mainBib}% Produces the bibliography via BibTeX.